\documentclass[runningheads]{llncs}
\usepackage[T1]{fontenc}
\usepackage{amsmath}
\usepackage{newtxtext}       %
\usepackage[varvw]{newtxmath}       

\usepackage{graphicx}
\usepackage{wrapfig}
\usepackage{times}
\usepackage{xspace}
\usepackage{graphicx}
\usepackage{dsfont}
\usepackage{hyperref}
\usepackage[normalem]{ulem} 
\newcommand{\G}{{\cal G}}
\newcommand{\tG}{\tilde{G}}

\renewcommand{\L}{{\cal L}}

\newcommand{\out}{\text{out}}
\newcommand{\play}{\text{play}}
\renewcommand{\path}{\text{path}}
\newcommand{\val}{\text{val}}
\newcommand{\util}{\text{util}}

\newcommand{\PO}{Player~$1$\xspace}
\newcommand{\PT}{Player~$2$\xspace}

\newcommand{\PLi}{Player~$i$\xspace}

\usepackage[dvipsnames]{xcolor}

\newcommand{\Nat}{\mathds{N}}

\newcommand{\zug}[1]{\langle #1  \rangle}
\newcommand{\set}[1]{\{ #1 \}}

\newcommand{\stam}[1]{}
\newcommand{\short}[1]{}

\usepackage{tikz,pgfplots}
\usetikzlibrary{automata, arrows.meta, positioning,decorations.markings, decorations.pathmorphing, shapes.misc}

\def\mathunderline#1#2{\color{#1}\underline{{\color{black}#2}}\color{black}}
\def\mathoverline#1#2{\color{#1}\overline{{\color{black}#2}}\color{black}}

\begin{document}
\title{Playing Against a Partially-Known Opponent}
%
%
\author{Eden Asta \and Guy Avni} 
%
\authorrunning{anonymous}
%
\institute{Department of Computer Science, University of Haifa}
\maketitle              
\begin{abstract}
A primary application of two-player graph games is {\em reactive synthesis}, in which the goal is to design a {\em reactive} system that ensures the specification in any environment. Synthesis traditionally takes a worst-case approach: we seek a strategy for the system player that wins against an adversarial environment player. 
In many situations, however, some information about the environment is available. 
We introduce and study {\em color-restricted adversary} (CRA) games that model environments with an agent who follows a strategy (color) chosen from a collection of previously known strategies. For example, the environment of a warehouse robot consists of another delivery robot who we know chooses a shortest route to a package, but we do not know which package it is targeting. 
 We first study an adversarial choice of colors and show that deciding the winner is NP-complete, which is surprising given that CRA games are exponentially-succinct turn-based games. 
 We consider CRA games in which colors are selected according to a known probability distribution and reduce the problem of finding the value of the game to an integer linear program. We study the problem of finding an optimal mixed strategy for the proponent as well as variants of CRAs. 

\keywords{Graph games \and parity games \and partial-information games}
\end{abstract}
\section{Introduction}\label{intro}
Two-player graph games constitute a fundamental model  with applications, e.g., in multi-agent systems~\cite{AHK02}, and a deep connection to foundations of logic~\cite{Rab69}. 
A primary application of graph games is {\em reactive synthesis}~\cite{PR89}, where the goal is to decide whether, given LTL specification $\varphi$, there is an implementation of a system such that no matter how the environment acts, the interaction produces a trace that satisfies $\varphi$. Synthesis reduces to a two-player graph game in which \PO is associated with the system, and his goal is to satisfy $\varphi$ no matter how \PT (the environment) chooses the uncontrollable actions. A {\em winning} \PO strategy is output as a system implementation.

\begin{wrapfigure}[11]{r}{0.2\textwidth}
\centering
\includegraphics[height=2cm]{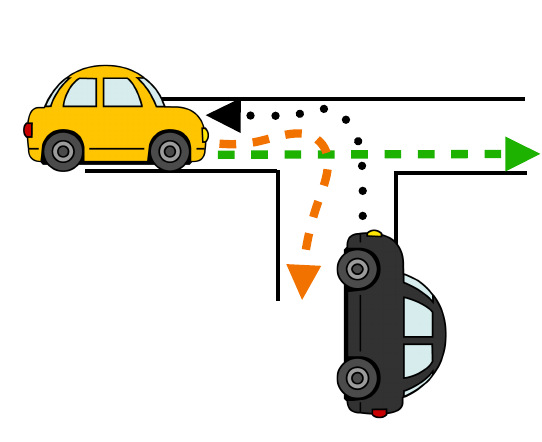}
\caption{The black vehicle needs to cross safely.}
\label{fig:cars}
\end{wrapfigure}
Traditionally, no prior knowledge is assumed on the environment, thus \PT is modeled  as purely adversarial. 
We consider environments with an agent who follows a strategy chosen from a collection of previously known strategies, as we demonstrate below. 




\smallskip\noindent{\bf Autonomous vehicle.}  Our task is to design a controller for the black vehicle in Fig.~\ref{fig:cars} 
whose specification is to turn left while avoiding 
the incoming yellow vehicle. Note that the yellow vehicle is either continuing: (1)~{\color{Green} straight}, meaning that the black vehicle needs to wait until the intersection is free, (2)~{\color{Orange} right}, meaning that the intersection is free and the black vehicle can enter immediately. We seek a controller for the black vehicle that crosses the intersection safely and as swiftly as possible and operates under the assumption that the environment (the yellow vehicle) chooses one of the available paths. 


\smallskip
\noindent{\bf Warehouse robots.} Consider a storage warehouse in which a robot $R_1$ delivers packages; it travels in the warehouse to its destination package, collects it, and delivers it to a drop-off location. Successful operation of the warehouse draws customers and increases the load on $R_1$. To ease waiting times, a second delivery robot $R_2$ is introduced. Our task is to synthesize a controller for $R_2$ for the specification: deliver packages while avoiding collision with $R_1$. Observe that $R_2$'s environment consists of $R_1$. While an adversarial $R_1$ would try to actively collide with $R_2$, this is clearly not how it operates and thus constitutes a crude abstraction. The owner, who now designs $R_2$, could have plenty of useful information regarding $R_1$; for example, the owner might know how $R_1$ is designed. While this is useful information, it does not make the task of synthesizing $R_2$ trivial. Indeed, observe that even if $R_2$ knows how $R_1$ operates, e.g., it takes a direct path to the target package, $R_2$ does not know which package is targeted. 




\paragraph*{Graph games with a color-restricted adversary.}
We introduce and study {\em color-restricted adversary} games (CRA games for short) that  can accommodate environment assumptions like the ones described above. Before defining CRA games, we recall traditional graph games. A graph game is played on a graph whose vertices are partitioned $V = V_1 \cup V_2$, where $V_i$ are the vertices that \PLi controls, for $i \in \set{1,2}$. The game proceeds as follows. A token is placed on some vertex. When the token is placed on vertex $v \in V_i$, \PLi chooses an outgoing edge from $v$ and moves the token across it. The interaction generates an infinite path, which determines the winner of the game. 

A CRA game is equipped with a set of colors $\set{1,\ldots, k}$ and a coloring function $l$ maps each outgoing edge from \PT vertices to a subset of colors, where each vertex has a unique outgoing edge colored with each color. The game proceeds as follows. Before the game starts, \PT is assigned a color $i \in \set{1,\ldots, k}$, and can only choose edges labeled with $i$. \PO knows the colors of the edges, but does not know \PT's color. \PO wins if he can guarantee his objective no matter which color \PT is following. 
Throughout the paper, we often consider an equivalent semantics: instead of fixing a color before the game starts, \PT chooses her color as the game progresses; she maintains a collection of colors that are consistent with her past choices, and in her turn, she adversarially chooses an edge that is consistent with her choice of colors while possibly revealing colors that she will not choose in the future. 
We illustrate the model.

\begin{figure}[htbp]
\begin{minipage}{0.5\textwidth}
\centering
\begin{tikzpicture}[
    roundnode/.style={circle, draw=black!60, fill=white!5, very thick, minimum size=5mm},
    smallround/.style={circle, draw=black!60, fill=white!5, very thick, minimum size=3.5mm},
    squarednode/.style={rectangle, draw=black!60, fill=white!5, very thick, minimum size=5mm},
]
    \draw (0,0) node[roundnode] (0) {$v_0$};
    \draw (2,0) node[squarednode] (1) {$v_1$};

    \draw (-2,0.6) node[squarednode] (2) {$v_2$};
    \draw (-2,-0.6) node[squarednode] (3) {$v_3$};

    \draw (2,0.8) node[smallround] (blue) {};
    \draw (2,-0.8) node[smallround] (red) {};

    \draw (-4,0.6) node[roundnode] (s) {$s$};
    \draw (-4,-0.6) node[roundnode, double] (t) {$t$};

    \draw (0,-0.6) edge[->,draw=black!60, very thick] node[] {} (0);
    \filldraw[black] (0,-0.6) circle (2pt) node[]{};

    \draw (0) edge[->,draw=black!60, very thick] node[] {} (1);
    \draw (0) edge[->,draw=black!60, very thick] node[] {} (2);
    \draw (0) edge[->,draw=black!60, very thick] node[] {} (3);

    \draw (blue) edge[->,draw=black!60, very thick] node[] {} (0);
    \draw (red) edge[->,draw=black!60, very thick] node[] {} (0);

    \draw (1) edge[->,draw=red, very thick, dotted] node[] {} (red);
    \draw (1) edge[->,draw=blue, very thick, densely dashed] node[] {} (blue);

    \draw (2) edge[->,draw=blue, very thick, dashed] node[] {} (s);
    \draw (2) edge[->,draw=red, very thick, loosely dotted] node[] {} (t);

    \draw (3) edge[->,draw=red, very thick, loosely dotted] node[] {} (s);
    \draw (3) edge[->,draw=blue, very thick, dashed] node[] {} (t);

\end{tikzpicture}

	\caption{A reachability CRA game in which \PO requires a memory to win.}\label{fig:ex1}
\end{minipage}
\hspace{0.4cm}
\begin{minipage}{0.5\linewidth}
	\centering
	\begin{tikzpicture}[
        roundnode/.style={circle, draw=black!60, fill=white!5, very thick, minimum size=5mm},smallround/.style={circle, draw=black!60, fill=white!5, very thick, minimum size=3.5mm},
        squarednode/.style={rectangle, draw=black!60, fill=white!5, very thick, minimum size=5mm},
        ]
		\draw (0,0) node[roundnode] (0) {$v_0$};

		\draw (0,1) node[squarednode] (1) {$v_1$};

        \draw (0,-1) node[squarednode] (2) {$v_2$};

		\draw (-2,0) node[roundnode, double] (3) {$v_3$};

        \draw (2,0) node[roundnode, double] (4) {$v_4$};

        \draw (-1,0) node[smallround] (t1) {${\color{red}t_1}$};

        \draw (1,0) node[smallround] (t2) {${\color{blue}t_2}$};

        \draw (0.6,-0.6) edge[->,draw=black!60, very thick, minimum size=7mm] node[] {} (0);

        \filldraw[black] (0.6,-0.6) circle (2pt) node[]{};

		\draw (0) edge[->,draw=black!60, very thick, minimum size=7mm] node[] {} (1);
        \draw (0) edge[->,draw=black!60, very thick, minimum size=7mm] node[] {} (2);
        \draw (3) edge[->,draw=black!60, very thick, minimum size=7mm] node[] {} (t1);
        \draw (4) edge[->,draw=black!60, very thick, minimum size=7mm] node[] {} (t2);

        \draw (2) edge[bend left,->,draw=red, very thick, loosely dotted, minimum size=7mm] node[] {} (t1);
        \draw (1) edge[bend right,->,draw=red, very thick, loosely dotted, minimum size=7mm] node[] {} (3);

        \draw (1) edge[bend left,->,draw=blue, very thick, dashed, minimum size=7mm] node[] {} (t2);
        \draw (2) edge[bend right,->,draw=blue, very thick, dashed, minimum size=7mm] node[] {} (4);

	\end{tikzpicture}
	\caption{A CRA game in which \PO loses, but as a uniform p-CRA or using a mixed strategy, \PO can win with probability $0.5$.}\label{fig:ex2}
\end{minipage}
\end{figure}
\begin{example}\label{ex:non-memoryless}
    Consider the game depicted in Fig.~\ref{fig:ex1}. 
\footnote{We stress that edges of a CRA are labeled with a subset of colors, but for sake of presentation, we often depict these as multi-edges in figures.} 
    The initial vertex is $v_0$ and \PO's goal is to reach $t$ (\PT wins if $s$ is reached). \PT follows either a {\color{Red} \dotuline{Red}} or {\color{blue} \dashuline{Blue}} strategy. 
    We claim that \PO wins from $v_0$. Observe first that \PO cannot win by proceeding directly to the target; indeed, if he moves to $v_2$, \PT chooses color {\color{blue} \dashuline{Blue}} and proceeds to $s$, and dually, \PT chooses color {\color{Red} \dotuline{Red}} if \PO proceeds to $v_3$. Instead, a winning \PO strategy proceeds first to $v_1$. \PT is then forced to reveal her true color; if she proceeds up it is {\color{blue} \dashuline{Blue}} and {\color{Red} \dotuline{Red}} otherwise. Now, at $v_0$, \PO knows how to win; proceed to $v_3$ if \PT is {\color{blue} \dashuline{Blue}} and to $v_2$ otherwise. 
    We note that this \PO winning strategy is not {\em memoryless}. Indeed, observe that at $v_0$, \PO first proceeds right, then proceeds left. In fact, in the second visit to $v_0$, \PO's choice depends on \PT's previous choice at $v_1$. 
 It is not hard to see that \PO has no winning memoryless strategy. 
\stam{    
     but there does not exist a winning memoryless strategy for Player 1 from $v_0$.
    The only edges leading to $t$ are from the vertexes $v_2, v_3$, however there are also edges leading to $s$ leaving $v_2, v_3$. 
    Since the edges $\zug{v_2,s}, \zug{v_3,t}$ belong to Player 2's blue strategy, and the edges $\zug{v_2,t}, \zug{v_3,s}$ belong to Player 2's red strategy, Player 1 cannot force the game into the target $t$ without knowing Player 2's real strategy.
    If in the first move Player 1 chooses to go to $v_1$, 
    In a memoryless strategy Player 1 would have to choose a single edge from the set $\{\zug{v_0,v_2}, \zug{v_0,v_2}, \zug{v_0,v_1}\}$ to always go to from $v_0$, neither of these three options can guaranty that Player 1 would win.
    }
\end{example}


Controller synthesis with an environment assumption like in the vehicle or robot examples above naturally reduce to solving a CRA game: the environment has a collection of memoryless strategies (each with its separate color) from which it chooses one adversarially, where each strategy maps a vertex to a neighbor. The examples demonstrate natural settings that give rise to such assumptions: we know how an agent in the environment operates to achieve its goal as well as the possible goal, but we do not know which specific goal they are targeting. 



\paragraph{Adversarial choice of colors.}
We study the problem of deciding whether \PO wins a parity CRA game against an adversarial choice of colors. 
We note that CRA games are succinctly-represented turn-based games; 
a game with $n$ vertices and $k$ colors has $n \cdot 2^k$ underlying {\em configurations}, where each configuration marks the subset of colors that are consistent with \PT's choices so far. Thus, a naive solution to a CRA game is in EXPTIME. We show that, somewhat surprisingly, the problem is NP-complete. 
The improved upper bound is obtained by establishing a linear upper bound on the number of configurations that are reachable for a fixed \PO strategy.


\paragraph{Probabilistic choice of colors.}
Often the choice of strategies follow patterns on which the system designer could have statistics. For example, in the vehicle example, proceeding straight could be more likely if that road is a highway, and in the robot example, it is often the case that the frequency at which packages are requested is known. 
We incorporate such assumptions by studying CRA games that are accompanied by a probability distribution over colors, called p-CRAs. \PO's goal in these game is to choose a strategy that maximizes the probability of winning. We show that the problem of finding an optimal strategy reduces to a polynomial-sized integer linear program. We stress that the strategies that we find in p-CRA games are history dependent. 
\begin{example}
\label{ex:pUCRA}
Consider the game depicted in Fig.~\ref{fig:ex2}. \PO needs to  either reach $v_3$ or $v_4$. Observe that he loses: if he chooses $v_1$, \PT chooses {\color{blue} \dashuline{Blue}} and proceed to $t_2$, and if \PO chooses $v_2$, \PT chooses {\color{Red} \dotuline{Red}} and proceeds to $t_1$. Suppose now that \PT is in fact choosing a path to a target, either target $t_1$ or $t_2$. 
Suppose that each target is chosen with uniform probability; \PO knows this probability distribution but not the choice of target. Then, proceeding to $v_1$ guarantees winning with probability $0.5$; indeed, \PO wins against a {\color{Red} \dotuline{Red}} opponent and loses against a {\color{blue} \dashuline{Blue}} opponent.
\end{example}

\paragraph{Mixed strategies.}
We study CRA games in which \PO is allowed to choose {\em mixed} strategies, namely strategies that assign a probability distribution over successor. Allowing mixed strategies is common practice in game theory (e.g.,~\cite{Neu28}) and it is not hard to show examples in which it is useful in UCRA games: Example~\ref{ex:pUCRA} shows that \PO has no winning strategy in the game depicted in Fig.~\ref{fig:ex2}, however, a strategy that uniformly at random chooses a successor at $v_0$ guarantees winning with probability $0.5$. 
We develop a PSPACE algorithm to find an optimal \PO mixed strategy in a CRA game. The algorithm scans the possible configurations of the game, but now, the recursive step involves constructing a matrix game and finding a max-min optimal strategy in the game. 

\paragraph{Variants.} Finally, we consider variants of CRA games. (1)~{\em Color restart.} We consider a game that includes edges that, when crossed, restart \PT's available choice of colors. The motivation arises from the robot example: we expect $R_2$ to target a new package (color) once a package is dropped. We show that the extended model retains the NP-complete bounds for reachability objectives. (2)~{\em Multi-player}. We consider a multi-player version of the game in which \PO plays against multiple color-restricted opponents. This corresponds to robot $R_2$ being introduced to a warehouse with multiple robots, not only a single one. (3)~{\em Generalized CRA games.} We allow multiple edges with the same color. This corresponds to a setting in which even if $R_2$ knows $R_1$ target package, there could be several paths to the target from which $R_1$ chooses. Such games are PSPACE-complete.

\paragraph*{Related work}
There has been extensive work that relaxes the adversarial assumption on the environment, specifically modeling environments that consist of agents as multi-player games, e.g.~\cite{AHK02,FKL10,CHJ08,MPS19}. CRA games offer a refined framework for incorporating natural environment assumptions while still modeling uncertainty as adversarial. CRA games are essentially {\em partial information} games, but an advantage of our refined modeling is we obtain improved complexity bounds; partial information games are EXPTIME-hard already for safety games~\cite{BD08}. There are technical similarities with works that study games in which parts of the arena are estimated, for example, energy estimates in energy games~\cite{KS25} or rewards in mean-payoff games~\cite{KPR18}. Unlike these works, CRA games incorporate behavioral prior knowledge on the opponent. Our assumption on \PT can be phrase a specific {\em multi-strategy} in the terminology of \cite{BDMR09}, namely a multi strategy whose set of actions (colors) decreases throughout the play as colors are ruled out. We stress that \cite{BDMR09} study a very different setting: their goal is to design (most) permissive multi strategies for their proponent, whereas, our proponent is playing against an opponent that is restricted to follow this specific type of multi strategy. 

\paragraph{Multi-environment MDPs.} 
A {\em Markov decision process} (MDP) can be seen as one-player game in which in each state, each action is associated with a probability distribution over neighboring states. 
A {\em multi-environment MDPs} (ME-MDPs)~\cite{RS14} is a collection of MDPs over the same state space and actions, but that differ in their transition functions (see the recent survey~\cite{Ras26}). CRAs can be seen as a special case of ME-MDPs in which the distributions are Dirac, i.e., for each distribution, state, and action, there is a unique successor. Similar to our study, ME-MDPs have been studied both under adversarial and probabilistic choice of transition function~\cite{CC+20,CD+25,BR26}. The complexity of solving ME-MDPs is typically complete for PSPACE or EXPTIME. 
CRAs constitute a natural fragment of ME-MDPs with improved complexity bounds and arguably cleaner and approachable definitions that amendable for generalizations, see our variants (restarting color and symmetric players), which to the best of our knowledge, have not been studied in ME-MDPs. 
We point to~\cite{SKD21} that finds an optimal policy in the probabilistic adversary model by a reduction to MILP, but the policy that they find is memoryless, i.e., it assigns an action for each state and time step, whereas the strategies that we construct via ILP in Sec.~\ref{sec:prob} are memoryfull, i.e., could assign a different action in a state depending on the history of choices made by \PT.


\stam{
Games played partially known arenas. 
\begin{itemize}
\item Energy games in which each edge is associated with an estimate on the true energy and the sum of energies in the arena is known. The true energy of an edge is either revealed when it is traversed or it is deduced when true energies are revealed by subtracting from the sum of energies. https://www.cs.huji.ac.il/~ornak/publications/atva25.pdf
\item Learning weights in mean-payoff: https://arxiv.org/abs/1804.08924
\item Nash equilibria in concurrent games: https://arxiv.org/abs/2207.02596
\end{itemize}
}

\section{Color-Restricted-Adversary Games}
A {\em color-restricted adversary (CRA)} game is played on a CRA arena $\zug{V, E, k, l}$, where both vertices $V=V_1\cup V_2$ and edges $E = E_1\cup E_2 \subseteq V\times V$ are partitioned among the players, the vertices that \PLi \ {\em controls} are $V_i$, and $E_i \subseteq V_i \times V$ are \PLi edges, for $i \in \set{1,2}$, the colors are $k \in \Nat$ and we denote $[k]=\{1,...,k\}$, and $l:E_2\to 2^{[k]}$ associates each \PT edge with a subset of colors. 
For a vertex $v\in V$, we denote the outgoing edges from $v$ by $\out(v)=\{e=\zug{v,u}\in E\}$ and for $v \in V_2$ and $j \in [k]$, the $j$-colored outgoing edges from $v$ by $\out_j(v)=\{e \in \out(v): j \in l(e)\}$. We assume that each vertex $v \in V_2$ has exactly one $j$-colored outgoing edge, for every color $j \in [k]$; namely, for all $v\in V_2$ and $j \in [k]$, we have $|\out_j(v)| =1$. Observe that this implies that for $e_1\neq e_2\in \out(v)$, we have $l(e_1)\cap l(e_2)=\emptyset$. 


\paragraph*{Objectives}
An outcome of a CRA game corresponds to a path in the graph $\zug{V, E}$, which determines the winner of the game. Formally, an objective is $O\subseteq V^\omega$, and we consider standard $\omega$-regular objectives:
\begin{description}
\item[Reachability.] \PO's {\em target} vertices are $T \subseteq V$ and a path $\pi \in V^\omega$ is winning iff $\pi$ visits $T$. 
\item[Parity (``max odd'').] Each vertex is associated with a parity index, given by $p: V \rightarrow \set{1,\ldots,d}$. 
For a path $\pi=v_0,v_1,...$, denote the vertices that $\pi$ visits infinity often by $\inf(\pi)$. \PO wins $\pi$ iff $\max\{ p(v):v\in \inf(\pi)\}$ is odd. 
\end{description}

\paragraph*{Semantics via the explicit game}
Consider a CRA parity game $\G = \zug{V, E, k, l, p}$. We describe the semantics of $\G$ by constructing the explicit turn-based game $\tG$ that corresponds to $\G$. See Fig.~\ref{fig:ex1exp} in the appendix for an example. 

Let $\tG = \zug{C, A, p}$. The vertices $C$ of $\tG$ are the {\em configurations} of $\G$, where a configuration is $c=\zug{v, L} \in C$, means that the token is placed on $v$ and $L \subseteq [k]$ are the \PT colors that have {\em not} been ruled out yet. Observe that $|C| = |V| \cdot 2^k$, namely it is exponential in $\G$. 
The player that controls $v$ in $\G$ controls $c$ in $\tG$, thus $C_1 = \set{\zug{v, L} \in C: v \in V_1}$ and $C_2 = \set{\zug{v, L} \in C: v \in V_2}$. 
Let $v_0 \in V$ be an initial vertex, then the initial configuration is $\zug{v_0, [k]}$. 
We describe the edges $A$ of $\tG$ by specifying the players' allowed actions and defining which configuration they lead to. 
Consider a configuration $c = \zug{v, L}$. If $v \in V_1$, \PO is allowed to move the token across any edge $e = \zug{v, v'} \in \out(v)$, resulting in the next configuration $c' = \zug{v', L}$. That is, no colors are ruled out when \PO acts. On the other hand, if $v \in V_2$, then \PT is only allowed to choose edges that are labeled with colors that have not been ruled out, i.e., an edge $e \in \out_j(v)$ is allowed iff $j \in L$. Choosing $e = \zug{v, v'}$ possibly rules out colors. The next configuration $c'$ is $\zug{v', L'}$ with $L' = L\cap l(e)$. That is, every color in $L\setminus l(e)$ is ruled out. We call $c'$ a {\em neighboring} configuration of $c$ and denote by $N(c)$ the neighbors of $c$.


\noindent{\bf Copies of $\G$.}
 Let $L \subseteq [k]$. The configurations in which the remaining possible \PT colors are $L$ form a {\em copy} of $\G$ in $\tG$ denoted $\tG_L=\zug{V_L,E_L}$, where $V_L = \set{c\in C : c=\zug{v,L}}$ and $E_L = \set{\zug{c,c'}: c,c' \in V_L \text{ and } c' \in N(c)}$. Intuitively, an edge within a copy does not rule out colors, thus does not expand our knowledge, whereas an edge between two different copies rules out at least one color, and so, expands our knowledge.
The {\em level} of $\tG_L$ is $|L|$. Note that edges out of copy $\tG_L$ lead to a copy with lower level. 

\noindent{\bf Strategies.}
For $i \in \set{1,2}$, a \PLi \ {\em strategy} is $\sigma_i : V_i\times 2^{[k]} \to E$. Note that $\sigma_i$ is {\em memoryless} in $\tG$, and this suffices for the objectives that we consider. 
A pair of strategies $\sigma_1$ and $\sigma_2$ for the two players and an initial configuration $c_0$, gives rise to a unique {\em play} denoted by $\play(c_0, \sigma_1, \sigma_2)$, which is defined inductively, where the inductive step, namely the definition of how a configuration is updated given an actions from the relevant players strategy, has already been described above.
Let $\play(c_0, \sigma_1, \sigma_2)=c_0,c_1,\ldots$, where $c_i = \zug{v_i,L_i}$.
The {\em path} that corresponds to $\play(c_0, \sigma_1, \sigma_2)$ is $\path(c_0, \sigma_1, \sigma_2) = v_0, v_1, \ldots$.

A \PO strategy $\sigma_1$ is {\em winning} from a configuration $c_0$ iff for any \PT strategy $\sigma_2$, we have $\path(c_0, \sigma_1, \sigma_2)\in O$.
Dually, a \PT strategy $\sigma_2$ is winning iff for any \PO strategy
$\sigma_1$, we have $\path(c_0, \sigma_1, \sigma_2)\notin O$.
For $i \in \{1, 2\}$,
a \PLi strategy $\sigma_i$ is winning strategy from vertex $v_0$ iff it is a winning strategy from the configuration $\zug {v_0, [k]}$.
We say that Player $i$ wins from a configuration $c_0$ if he has a winning strategy from $c_0$, and wins from $v_0\in V$ if he has a a wining strategy from $v_0$.

\section{Playing Against an Adversarial Choice of Colors}
In this section, we prove the following theorem:

\begin{theorem}
Deciding whether \PO has a winning strategy from an initial vertex is NP-complete for CRA games.
Moreover, hardness applies for reachability objectives and the upper bounds for parity objectives. 
\end{theorem}

We start with the lower bound. 

\begin{lemma}
\label{lem:UCRA-NP-hard}
Deciding whether \PO wins a CRA reachability game is NP-hard.
\end{lemma}
\begin{proof} 
We show a reduction from 3SAT. 
Consider a formula $\varphi=C_1\wedge \ldots \wedge C_k$ over variables $x_1,\ldots,x_m$.
We construct a CRA reachability game $\G = \zug{V, E, k, l ,T}$. 
See an example in Fig.~\ref{fig:multoptpayoff}.
Let $V_1 = \set{s_1,\ldots, s_m}\cup\{s\}$, $V_2 = \set{x_i, \overline{x_i}: 1 \leq i \leq m}$, and \PO's targets $T = \set{C_1,\ldots, C_k}$ are sinks.  
For $1 \leq i < m$, the neighbors of $s_i$ are $x_i$ and $\overline{x_i}$. Each clause is associated with a color. For $1 \leq i \leq m$,  $\ell_i \in \set{x_i, \overline{x_i}}$, and clause $C_j$, we define that the $C_j$-labeled edge from $\ell_i$ proceeds to target vertex $C_j \in T$ if $\ell_i$ appears in $C_j$, and otherwise it proceeds to $s_{i+1}$, or $s$ in the case that $i=m$. 
For example, observe Fig.~\ref{fig:multoptpayoff}; setting $f_1(x_1) = 1$ satisfies {\color{red} \dotuline{$C_1$}} and {\color{blue} \dashuline{$C_2$}} thus the corresponding {\color{red} \dotuline{Red}} and {\color{blue} \dashuline{Blue}} edges lead from from $x_1$ to $T$ whereas from $\overline{x_1}$, they lead to $s_2$. 
Vertex $s$ is a sink and \PT wins if the game reaches $s$. 

Intuitively, \PO chooses an assignment $f$ to the variables, one variable at a time; when he moves from $s_i$ to $x_i$ or $\overline{x_i}$ he assigns, respectively, $1$ or $0$ to the variable $x_i$. For $1 \leq i \leq m$, let $f\big|_{i}: \set{x_1,\ldots, x_{i}} \rightarrow \set{1,0}$ denote \PO's choice of partial assignment by turn $i$. Let $S_{i} \subseteq [k]$ denote the clauses that $f\big|_{i}$ satisfies. That is, $C_j \in S_i$ iff there exists a literal $\ell$ in $C_j$ such that $f\big|_{i}(\ell) = 1$. Suppose that the game reaches $\ell_i \in \set{x_i, \overline{x_i}}$. It is \PT's turn. Observe that clauses that have already been satisfied, namely clauses in $S_{i-1}$, correspond to colors that are inconsistent with \PT's previous choices, and are thus disallowed. Further observe that each clause $C_j \in (S_i \setminus S_{i-1})$ leads to $T$ and results in \PT losing, and for $C_j \notin S_i$, the edge colored $C_j$ leads to $s_{i+1}$. We thus assume that if possible, \PT proceeds to $s_{i+1}$. 

We claim that \PO wins iff $\varphi$ is satisfiable. First, if there is a satisfying assignment $f$ to $\varphi$, \PO wins by choosing $f$. Indeed, since $f$ satisfies all clauses, \PT will eventually be forced to choose a color leading to a target. Second, if $\varphi$ is not satisfiable, for every partial assignment $f\big|_{i}$, for $1 \leq i \leq m$, there is at least one clause $C_j$ that is not satisfied, thus \PT can choose the edge leading to $s_{i+1}$ and eventually reach $s$.

\begin{figure}

	\centering

	\begin{tikzpicture}[
        roundnode/.style={circle, draw=black!60, fill=white!5, very thick, minimum size=5mm},smallround/.style={circle, draw=black!60, fill=white!5, very thick, minimum size=3.5mm},
        squarednode/.style={rectangle, draw=black!60, fill=white!5, very thick, minimum size=5mm},
        tightzigzag/.style={decorate, decoration={zigzag, segment length=1.5mm, amplitude=.5mm}}
        ]

        \draw (-6,0) node[roundnode] (s1) {$s_1$};		
        \draw (-2,0) node[roundnode] (s2) {$s_2$};
        \draw (2,0) node[roundnode] (s3) {$s_3$};
        \draw (6,0) node[roundnode] (s4) {$s$};

        \draw (-4.5,0.5) node[squarednode] (x1) {$x_1$};
        \draw (-4.5,-0.5) node[squarednode] (nx1) {$\overline{x_1}$};

        \draw (-0.5,0.5) node[squarednode] (x2) {$x_2$};
        \draw (-0.5,-0.5) node[squarednode] (nx2) {$\overline{x_2}$};

        \draw (3.5,0.5) node[squarednode] (x3) {$x_3$};
        \draw (3.5,-0.5) node[squarednode] (nx3) {$\overline{x_3}$};

        \draw (-3.75,-2) node[roundnode, double, minimum size=7mm] (c1) {$C_1$};
        \draw (-0.25,-2) node[roundnode, double, minimum size=7mm] (c2) {$C_2$};
        \draw (3.25,-2) node[roundnode, double, minimum size=7mm] (c3) {$C_3$};

        \draw (-7,0) edge[->,draw=black!60, very thick, minimum size=7mm] node[left] {} (s1);

        \filldraw[black] (-7,0) circle (2pt) node[]{};

		\draw (s1) edge[->,draw=black!60, very thick, minimum size=7mm] node[] {} (x1);
        \draw (s1) edge[->,draw=black!60, very thick, minimum size=7mm] node[] {} (nx1);
        
        \draw (s2) edge[->,draw=black!60, very thick, minimum size=7mm] node[] {} (x2);
        \draw (s2) edge[->,draw=black!60, very thick, minimum size=7mm] node[] {} (nx2);

        \draw (s3) edge[->,draw=black!60, very thick, minimum size=7mm] node[] {} (x3);
        \draw (s3) edge[->,draw=black!60, very thick, minimum size=7mm] node[] {} (nx3);

        \draw (x1) edge[out=0, in=-270, looseness=0.5,->,draw=red, very thick, dotted, minimum size=7mm] node[] {} (c1);
        \draw (nx1) edge[bend right,->,draw=red, very thick, dotted, minimum size=7mm] node[] {} (s2);

        \draw (x2) edge[out=-20, in=20, in looseness=1.5, out looseness=2, ->,draw=red, very thick, dotted, minimum size=7mm] node[] {} (c1);
        \draw (nx2) edge[bend right, ->,draw=red, very thick, dotted, minimum size=7mm] node[] {} (s3);

        \draw (x3) edge[out=-30, in=15, in looseness=1.3, out looseness=1.2, ->,draw=red, very thick, dotted, minimum size=7mm] node[] {} (c1);
        \draw (nx3) edge[->,draw=red, very thick, dotted, minimum size=7mm] node[] {} (s4);

        \draw (x1) edge[out=0, in=-210, out looseness=0.5, in looseness=2,->,draw=blue, very thick, dashed, minimum size=7mm] node[] {} (c2);
        \draw (nx1) edge[->,draw=blue, very thick, dashed, minimum size=7mm] node[] {} (s2);

        \draw (x2) edge[->,draw=blue, very thick, dashed, minimum size=7mm] node[] {} (s3);
        \draw (nx2) edge[out=0, in=80, in looseness=1, out looseness=1, ->,draw=blue, very thick, dashed, minimum size=7mm] node[] {} (c2);

        \draw (x3) edge[out=0, in=10, in looseness=3, out looseness=0.4, ->,draw=blue, very thick, dashed, minimum size=7mm] node[] {} (c2);
        \draw (nx3) edge[bend right, ->,draw=blue, very thick, dashed, minimum size=7mm] node[] {} (s4);

        \draw (x1) edge[->,draw=green, thick, tightzigzag, minimum size=7mm] node[] {} (s2);
        \draw (nx1) edge[->,draw=green, thick, tightzigzag, minimum size=7mm] node[] {} (c3);

        \draw (x2) edge[out=0, in=120, out looseness=0.5, in looseness=0.3, ->,draw=green, thick, tightzigzag, minimum size=7mm] node[] {} (c3);
        \draw (nx2) edge[->,draw=green, thick, tightzigzag, minimum size=7mm] node[] {} (s3);

        \draw (x3) edge[->,draw=green, thick, tightzigzag, minimum size=7mm] node[] {} (s4);
        \draw (nx3) edge[out=0, in=50, in looseness=1, out looseness=1, ->,draw=green, thick, tightzigzag, minimum size=7mm] node[] {} (c3);

	\end{tikzpicture}

	\caption{The CRA reachability game constructed from the formula $\varphi = C_1 \wedge C_2 \wedge C_3$ with $\color{red}{\dotuline{C_1}}\color{black}=(x_1\vee x_2\vee x_3)$, $\color{blue}{\dashuline{C_2}}\color{black}=(x_1\vee \overline{x_2}\vee x_3)$, and $\color{green!60!black}{\uwave{C_3}}\color{black}=(\overline{x_1}\vee x_2\vee\overline{x_3})$.}\label{fig:multoptpayoff}
\end{figure}
\end{proof}

\subsection{Upper bound}
\label{sec:NP-upper}
For the remainder of the section, fix a CRA game $\G = \zug{V, E, k, l, p}$ and an initial vertex $v_0 \in V$. 
Before presenting the NP upper bound, we present a warm-up PSPACE algorithm: 

\paragraph{Warm-up upper bound.}
\stam{
\begin{lemma}
\label{lem:CRA-PSPACE}
Given a CRA parity game $\G = \zug{V, E, k, l, p}$ and an initial vertex $v_0$ deciding whether Player 1 wins from $v_0$ is in PSPACE. 
\end{lemma}
}
Recall that $\tG$ contains copies of $\G$. For a collection of colors $L \subseteq [k]$, the copy $\tG_L$ corresponds to configurations in which \PT's choices are consistent with the colors in $L$. In $\tG_L$, \PT can choose any edge $e$ with $l(e) \cap L \neq \emptyset$. The game stays in $\tG_L$ if \PT chooses an edge $e$ with  $l(e) \subseteq L$, otherwise \PT reveals some colors that she is not following, and the game proceeds to copy $L \setminus l(e)$.

\smallskip
\noindent{\bf Observation:} 
A play eventually gets stuck in a single copy. Formally, for every play $\pi$ and two configurations $c = \zug{v, L}$ and $c' = \zug{v', L'}$ that $\pi$ traverses infinitely often, i.e., $c, c' \in \inf(\pi)$, we have $L=L'$.

\stam{
{\em Proof.} There are $k$ levels and $2^n$ copies of $G$ in $\tilde{G}$.
In each copy there are $n$ vertices - one for each $v\in V$, all with the same $L\subseteq [k]$.
Every edge in $\tilde{G}$ is either inside a copy, or from a copy $G_{L_1}$ to some copy $G_{L_2}$ in a deeper level since for every action $a=\zug{c_1,c_2}\in A$ and $c_1=\zug{v_1,L_1}$, $c_2=\zug{v_2,L_2}$ it follows that $L_2\subseteq L_1$.

There are no loop in $\tilde{G}$ that contain configurations in more than one copy. Formally for any set of configurations $c_0,c_1,...,c_l\in C$ such that for any $i\in\{0,...,l-1\}$  $\zug{c_i,c_j}\in A, c_i=\zug{v_i,L_i}$ and $c_0=c_l$ it follows that for any $L_i=L_j$, $i,j\in\{0,...,l\}$ . 
Every play on $\G$ would eventually get 'stuck' in a single copy $G_L$, that is to say for any path $\pi=c_0,c_1,...$ where $c_i=\zug{v_i,L_i}$, there exists $L\subseteq [k], i\ge 0$ such that for every $c_j\in inf(\pi),j\ge i$  it follows that  $L_j=L$. The lemma follows \hfill $\Box$

\medskip
}

\smallskip

The algorithm recursively traverses the graph of copies of $\tG$. 
By the observation, in a configuration $c_0 = \zug{v, L}$ in copy $\tG_L$, we decide whether \PO wins by accounting for plays that stay in $\tG_L$ and plays that exit it to a deeper copy. 
The base case is for singleton $L = \set{i}$, which corresponds to \PO knowing that \PT follows Color~$i$, thus \PT has no actions, and copy $\tG_L$ is in fact a graph with no outgoing edges. 
We describe a recursive call. Consider a copy $\tG_L$ for $|L| > 1$. Here, a \PT action corresponds to staying in the copy or exiting it by taking an edge $a = \zug{c, c'}$ that leads out of $\tG_L$. We recursively solve from $c'$, replace the sub-graph with a sink vertex whose parity depends on the winner in $c'$. 



\begin{definition}[The graph of reachable configurations]
Fix a \PO strategy $\sigma_1$. A configuration $c = \zug{v, L}$ is {\em reachable} if there exists a \PT strategy $\sigma_2$ such that $\play(v_0, \sigma_1, \sigma_2)$ visits $c$. 
For $L \subseteq [k]$, copy $L$ is {\em reachable} if there is a reachable configuration $\zug{v, L}$. 
The {\em graph of reachable configurations} $\tG[\sigma_1]$ is obtained from the explicit game $\tG = \zug{C, A, p}$ by omitting edges that do not agree with $\sigma_1$ and omitting unreachable vertices, thus $\tG[\sigma_1] =\zug{C',A'}$, where $C'$ are the reachable configurations and  $A'= A\setminus \{\zug{v,u}\in A: v \in V_1 \text{ and } \sigma_1(v)\neq u\}$. 
\end{definition}


The following lemma shows that $\G[\sigma_1]$ has a tree-like structure. 

\begin{lemma}
\label{lem:tree}
For every \PO strategy $\sigma_1$, the graph $\tG[\sigma_1]$ is a tree with cycles only in its bottom strongly-connected components. 
\end{lemma}
\begin{proof}
Consider a configuration $\zug{v, L}$ with $v \in V_2$. Recall that in CRA games, we have $\cup_{e \in \out(\zug{v,L})} l(e) = L$ and for $e \neq e' \in \out(\zug{v, L})$, we have $l(e) \cap l(e') = \emptyset$. Thus, there are two options: (1)~there is a unique outgoing edge $e = \out(\zug{v, L})$ that is associated with all colors $L$ and leads to a neighbor within copy $\tG_L$ or (2)~there are multiple outgoing edges, each edge rules out some colors, and thus leads to a different copy.  Further observe that since \PO's strategy is fixed, every \PO configuration has a unique neighbor and the neighbor is in the same copy. Thus, from each configuration in $\tG_L$ there is a unique path that either closes a loop in the copy (in which case this is a BSCC) or exits the copy. 

Assume towards contradiction that there is a configuration $\zug{v, L}$ that is reachable via two different simple paths $\pi_1$ and $\pi_2$. Let $L_1, L_2\ldots, L$ and $L'_1, L'_2, \ldots, L$ be the sequence of copies that the paths traverse. Recall that the sequences are monotonically decreasing, i.e., $L_1 \supseteq L_2 \supseteq \ldots \supseteq L$ and $L'_1 \supseteq L'_2 \supseteq \ldots \supseteq L$. Since both paths start from the same initial configuration, we have $L_1 = L'_1$. We argue that there is $j > 1$ such that $L_j \neq L'_j$. Indeed, as observed above, within each copy, the paths coincide, meaning that as long as the traversed copies coincide, the paths coincide. But we assume that $\pi_1$ differs from $\pi_2$. Thus, the paths fork at some \PT configuration $\zug{u, L_i}$: $\pi_1$ follows edge $e_1$ to $\zug{u_1, L_{i+1}}$ and $\pi_2$ follows edge $e_2$ to $\zug{u_2, L'_{i+1}}$ with $L_{i+1} \neq L'_{i+1}$. Recall that by traversing $e_1$, \PT reveals that she is not following colors $L_i \setminus L_{i+1}$. But since $\G$ is a CRA game, no color that is associated with $e_1$ is associated with $e_2$, meaning that $L \cap L'_{i+1} = \emptyset$ and contradicting that $\pi_2$ reaches $\zug{v, L}$. 
\end{proof}

Next, we bound the number of reachable configurations. 

\begin{lemma}
\label{lem:small-copies-tree}
For a \PO strategy $\sigma_1$, at most $2k-1$ copies are reachable in $\tG[\sigma_1]$.
\end{lemma}
\begin{proof}
A {\em copies tree} is a tree $T$ that intuitively could represent the structure of copies of $\tG[\sigma_1]$. Each vertex of $T$ is associated with a subset of colors $L \subseteq [k]$. The children of $L$ form a partition of $L$ with each part being non-empty: namely, let $L_1,\ldots, L_n$ be the children of $L$, then $L_i \neq \emptyset$, for all $1 \leq i \leq n$, we have $\cup_{1 \leq i \leq n} L_i = L$, and $L_i \cap L_j = \emptyset$, for $1 \leq i < j \leq n$. We call such a partition, an {\em $n$-partition of colors}. 
It is not hard to see that Lem.~\ref{lem:tree} implies that one could construct $T$ such that every path in $T$ corresponds to a sequence of copies that a path in $\tG[\sigma_1]$ traverses, omitting repetitions of copies. Note that the $|\tG[\sigma_1]| \leq |T| \cdot |\G|$. 
For $k$, we define $T_k$ to be the largest copies tree with $k$ colors. Thus, $|\tG[\sigma_1]| \leq |T_k| \cdot |\G|$. 

We bound the size of $T_k$. Suppose that the root of $T_k$ has $q$ children. For $1 \leq i \leq q$, denote by $L_i \subseteq [k]$ the colors that are associated with the $i$-th subtree and $a_i = |c_i|$ the number of colors. Since $T_k$ is largest, the $i$-th subtree is $T_{a_i}$, namely the largest tree with $a_i$ colors. Overall, the size of $T_k$ is the number of nodes in the subtrees in addition to the root, and we obtain $T_k$ by finding the $q$-partition of colors that achieves the largest size, thus $|T_k|=1+\max(\sum_{i=1}^q|T_{a_i}|)$. 

We claim that $T_k$ is obtained for $q = 2$, i.e., the root has two children. Assume towards contradiction that the root has $q>2$ children. We claim that one could replace the first $q-1$ children with a single child with $\sum_{i=1}^{q-1}a_i=k-a_q$ colors to obtain a larger tree. 
Indeed, observe that the following tree $T'$ is a legal copies tree: the root is associated with colors $\cup_{1 \leq i \leq q-1} c_i$ and its subtrees are the trees $T_{a_1},\ldots, T_{a_{q-1}}$. Observe $|T'| = 1 + |\sum_{1 \leq i \leq q-1} |T_{a_i}|$, and is in particular larger. Thus, by replacing the subtrees with $T'$, we obtain a strictly larger tree, in contradiction to $T_k$ being largest. 

We have narrowed down the structure of the root of $T_k$; it has two children. It remains to identify the maximal $2$-partition of $[k]$. We claim that all partitions lead to the same size of tree, namely for any $0<h_1, h_2<k$, we have $|T_{h_1}|+|T_{k-h_1}|=|T_{h_2}|+|T_{k-h_2}|$. 
The proof is by induction. For the base case $k=2$, and we have $|T_{1}|+|T_{1}|=|T_{1}|+|T_{1}|$.
Assume that the claim holds for $k'<k$ and we prove for $k$. 
As before, $|T_{h_2}| = |T_{h_1}|+|T_{h_2-h_1}|+1$ and $|T_{k-h_1}| = |T_{h_2-h_1}|+|T_{k-h_2}|+1$. 
Wlog, let $h_1<h_2$. 
Combining the two, we get $|T_{h_1}|+|T_{k-h_1}|={|T_{h_1}|+|T_{h_2-h_1}|}+|T_{k-h_2}|+1=|T_{h_2}|+|T_{k-h_2}|$.

Since all partitions lead to the same tree size, we use a partition that halves the colors, namely we assume that children of the root are $T_{\lfloor k/2\rfloor}$ and $T_{\lceil k/2\rceil}$. We conclude that $|T_k|=1+|T_{\lfloor k/2\rfloor}|+|T_{\lceil k/2\rceil}|$, for $k>1$, and  $|T_1|=1$. It is not hard to show that $|T_k|$ is the size of a full binary tree of height $\log(k)$, thus $|T_k|\leq2k-1$.
\end{proof}

We are ready to prove the complexity upper bound.

\begin{lemma}
\label{lem:UCRA-NP}
Deciding whether \PO wins a CRA parity game from initial vertex $v_0$ is in NP.
\end{lemma}
\begin{proof}
The algorithm guesses a copies tree $T$ and a strategy $\sigma_1$ only for the reachable copies of $T$. Note that by Lem.~\ref{lem:small-copies-tree}, the size of the copies tree is  $|T| \leq (2k-1)$ and thus the size of $\sigma_1$ is $|T| \cdot |\G|$, which is polynomial in the input. The algorithm then verifies the guess by scaning $\tG[\sigma_1]$ in a DFS manner, where in each copy $L$, if $L$ has children in $T$, one needs to verify that the unique path in $\tG_L$ leads to a \PT vertex whose children agree with the children of $L$ in $T$, and if $L$ is a leaf in $T$, one needs to verify that the unique path satisfies the parity objective. 
\end{proof}

We close this section by reasoning about strategy complexity. A finite-memory strategy is formally implemented as a {\em Moore machine} (see details, e.g., in~\cite{Fij26}). 
Lem.~\ref{lem:UCRA-NP} provides an upper bound on the implementation size of a strategy, namely $(2k-1) \cdot |V|$ states since the strategy switches between memoryless strategies, one for each reachable copy. 
We now show that this bound is tight.

We demonstrate the construction in Fig.~\ref{fig:prop11}, and the full proof is in App.~\ref{app:memory-req}. The idea generalizes Fig.~\ref{fig:ex1}. \PO wins by first proceeding right to $u^1_1$. \PT's action rules out either $\set{1,2}$ or $\set{3,4}$. Depending on \PT's action, \PO proceeds to either $u^1_2$ or $u^2_2$, to rule out the final color, and then proceeds left to win the game. Note that $2k-1$ copies are reachable. Indeed, a winning play visits $v_0$ three times; initially, in copy $\set{1,2,3,4}$, second, either in copy $\set{1,2}$ or $\set{3,4}$, and finally, in one of the copies $\set{1}, \set{2}, \set{3},$ or $\set{4}$, for a total of $2k-1=7$ copies. 

\begin{proposition}
\label{prop:memory-req}
For each $n\in \Nat$ and $k = 2^n$, there is a $k$-color CRA reachability game in which the number of reachable copies is $2k-1$. 
\end{proposition}

\begin{figure}[ht]
\centering
\begin{minipage}{0.5\textwidth}
	\begin{tikzpicture}[
        roundnode/.style={circle, draw=black!60, fill=white!5, very thick, minimum size=5mm},smallround/.style={circle, draw=black!60, fill=white!5, very thick, minimum size=3.5mm},
        squarednode/.style={rectangle, draw=black!60, fill=white!5, very thick, minimum size=5mm}, tightzigzag/.style={decorate, decoration={zigzag, segment length=1.5mm, amplitude=.5mm}}
        ]
		\draw (0,0) node[roundnode] (v0) {$v_0$};

        \draw (-1,1.5) node[squarednode] (v1) {$v_1$};

        \draw (-1,0.5) node[squarednode] (v2) {$v_2$};

		\draw (-1,-0.5) node[squarednode] (v3) {$v_3$};

        \draw (-1,-1.5) node[squarednode] (v4) {$v_4$};

        \draw (1.5,1.5) node[squarednode] (u11) {$u_1^1$};

        \draw (1.5,0) node[squarednode] (u21) {$u_2^1$};

        \draw (1.5,-1.5) node[squarednode] (u22) {$u_2^2$};

        \draw (-4,1.5) node[roundnode] (s) {$s$};

        \draw (-4,-1.5) node[roundnode, double] (t) {$t$};

        \draw (0,-1) edge[->,draw=black, very thick, minimum size=7mm] node[] {} (0);

        \filldraw[black] (0,-1) circle (2pt) node[]{};

		\draw (v0) edge[->,draw=black, very thick, minimum size=7mm] node[] {} (v1);
        \draw (v0) edge[->,draw=black, very thick, minimum size=7mm] node[] {} (v2);
        \draw (v0) edge[->,draw=black, very thick, minimum size=7mm] node[] {} (v3);
        \draw (v0) edge[->,draw=black, very thick, minimum size=7mm] node[] {} (v4);

        \draw (v0) edge[->,draw=black, very thick, minimum size=7mm] node[] {} (u11);
        \draw (v0) edge[->,draw=black, very thick, minimum size=7mm] node[] {} (u21);
        \draw (v0) edge[out=-50, in=180, looseness=0, ->,draw=black, very thick, minimum size=7mm] node[] {} (u22);

        \draw (u11) edge[out=-160, in=60, looseness=0.5, ->,draw=black!60, very thick, minimum size=7mm] node[midway, left, text=black] {\scriptsize $\{1,2\}$} (v0);

        \draw (u11) edge[out=-110, in=30, looseness=0.5, ->,draw=black!60, very thick, minimum size=7mm] node[midway, right, text=black] {\scriptsize $\{3,4\}$} (v0);

        \draw (u21) edge[out=160, in=15, looseness=0.3, ->,draw=black!60, very thick, minimum size=7mm] node[xshift=1.7mm, yshift=1mm, text=black] {\scriptsize $\{1\}$} (v0);

        \draw (u21) edge[out=-160, in=-15, looseness=0.3, ->,draw=black!60, very thick, minimum size=7mm] node[xshift=1.7mm, yshift=-1.5mm, text=black] {\scriptsize $\{2\}$} (v0);

        \draw (u22) edge[out=140, in=-30, looseness=0.4, ->,draw=black!60, very thick, minimum size=7mm] node[xshift=3mm, yshift=-1.5mm, text=black] {\scriptsize $\{3\}$} (v0);

        \draw (u22) edge[out=-160, in=-60, looseness=0.4, ->,draw=black!60, very thick, minimum size=7mm] node[xshift=-1.5mm, yshift=-2mm, text=black] {\scriptsize $\{4\}$} (v0);

        \draw (v1) edge[->,draw=black!60, very thick, minimum size=7mm] node[yshift=1.5mm, text=black!60] {\scriptsize $\{2,3,4\}$} (s);
        \draw (v2) edge[->,draw=black!60, very thick, minimum size=7mm] node[rotate=-15, xshift=0mm, yshift=1.5mm, text=black!60] {\scriptsize $\{1,3,4\}$} (s);
        \draw (v3) edge[->,draw=black!60, very thick, minimum size=7mm] node[ rotate=-30,xshift=-1.5mm, yshift=1.5mm, text=black!60] {\scriptsize $\{1,2,4\}$} (s);
        \draw (v4) edge[->,draw=black!60, very thick, minimum size=7mm] node[rotate=-45,xshift=-7mm, yshift=-1.7mm, text=black!60] {\scriptsize $\{1,2,3\}$} (s);

        \draw (v1) edge[->,draw=black, very thick, minimum size=7mm] node[rotate=40,xshift=-7mm, yshift=1mm, text=black] {\scriptsize $\{1\}$} (t);
        \draw (v2) edge[->,draw=black, very thick, minimum size=7mm] node[rotate=25,xshift=-2mm, yshift=-1.7mm, text=black] {\scriptsize $\{2\}$} (t);
        \draw (v3) edge[->,draw=black, very thick, minimum size=7mm] node[xshift=0mm, yshift=-1.5mm, text=black] {\scriptsize $\{3\}$} (t);
        \draw (v4) edge[->,draw=black, very thick, minimum size=7mm] node[yshift=-1.5mm, text=black] {\scriptsize $\{4\}$} (t);

	\end{tikzpicture}
	\caption{A CRA reachability game in which the number of reachable copies is $2 \cdot k -1$, for $k=4$. 
    }\label{fig:prop11}
\end{minipage}%
\begin{minipage}{0.5\textwidth}%
	\centering
	\begin{tikzpicture}[
        roundnode/.style={circle, draw=black!60, fill=white!5, very thick, minimum size=5mm},smallround/.style={circle, draw=black!60, fill=white!5, very thick, minimum size=3.5mm},
        squarednode/.style={rectangle, draw=black!60, fill=white!5, very thick, minimum size=5mm}, tightzigzag/.style={decorate, decoration={zigzag, segment length=1.5mm, amplitude=.5mm}}
        ]
		\draw (0,0) node[roundnode] (0) {$v_0$};

		\draw (1,0) node[squarednode] (3) {$v_3$};

        \draw (-1,0.5) node[squarednode] (1) {$v_1$};

		\draw (-1,-0.5) node[squarednode] (2) {$v_2$};

        \draw (1,1) node[squarednode] (4) {$v_4$};

        \draw (1,-1) node[squarednode] (5) {$v_5$};

        \draw (-2.25,0.5) node[roundnode] (s) {$s$};

        \draw (-2.25,-0.5) node[roundnode, double] (t) {$t$};

        \draw (0,-1) edge[->,draw=black!60, very thick, minimum size=7mm] node[] {} (0);

        \filldraw[black] (0,-1) circle (2pt) node[]{};

		\draw (0) edge[->,draw=black!60, very thick, minimum size=7mm] node[] {} (1);
        \draw (0) edge[->,draw=black!60, very thick, minimum size=7mm] node[] {} (2);
        \draw (0) edge[->,draw=black!60, very thick, minimum size=7mm] node[] {} (3);

        \draw (4) edge[->,draw=black!60, very thick, minimum size=7mm] node[] {} (0);
        \draw (5) edge[->,draw=black!60, very thick, minimum size=7mm] node[] {} (0);

        \draw (3) edge[->,draw=blue, very thick, dashed, minimum size=7mm] node[] {} (4);
        \draw (3) edge[->,draw=red, very thick, dotted, minimum size=7mm] node[] {} (5);
        \draw (3) edge[bend left,->,draw=green, thick, tightzigzag, minimum size=7mm] node[] {} (5);

        \draw (2) edge[->,draw=red, very thick, dotted, minimum size=7mm] node[] {} (s);
        \draw (2) edge[->,draw=blue, dashed, very thick, minimum size=7mm] node[] {} (t);
        \draw (2) edge[bend left,->,draw=green, thick, tightzigzag, minimum size=7mm] node[] {} (t);

        \draw (1) edge[->,draw=blue, very thick, dashed, minimum size=7mm] node[] {} (s);
        \draw (1) edge[->,draw=red, very thick, dotted, minimum size=7mm] node[] {} (t);
        \draw (1) edge[bend right,->,draw=green, thick, tightzigzag, minimum size=7mm] node[] {} (s);

	\end{tikzpicture}

	\caption{A probabilistic CRA reachability game.}\label{fig:p-UCRA}
\end{minipage}
\end{figure}


\section{Playing Against a Probabilistic Choice of Colors}
\label{sec:prob}
In the previous section, our goal was to find a \PO strategy that wins against an adversarial choice of colors. In this section, we assume that \PT's choice of color is random, and the goal is to maximize the probability of winning. We demonstrate with an example. 

\begin{example}
Consider the reachability CRA game that is depicted in Fig.~\ref{fig:p-UCRA}. Observe first that \PO cannot win. Indeed, similar to Example~\ref{ex:non-memoryless}, before proceeding left, \PO must pinpoint \PT's true color; if he proceeds to $v_1$ or $v_2$, \PT respectively chooses {\color{Green} \uwave{Green}} and {\color{Red} \dotuline{Red}} and the game proceeds to the losing sink $s$. Unlike Example~\ref{ex:non-memoryless}, by proceeding right, \PT does not reveal her true color, since proceeding down from $v_3$ to $v_5$ is consistent with both {\color{Green} \uwave{Green}} and {\color{Red} \dotuline{Red}}. 

Assume now, that \PT's color is chosen according to the distribution $\Pr({\color{red} \text{\dotuline{Red}}})=\frac{1}{2}, \Pr({\color{blue} \text{\dashuline{Blue}}})=\frac{1}{3}$, and $\Pr({\color{green!60!black}\text{\uwave{Green}}})=\frac{1}{6}$. Consider the following \PO strategy. He first proceeds right. If \PT's color is {\color{blue} \dashuline{Blue}}, the game proceeds to $v_4$, and \PO wins (with probability $1$) by proceeding from $v_0$ to $v_2$. Otherwise, \PT proceeds from $v_3$ to $v_5$ and \PO knows that \PT's color is either {\color{Green} \uwave{Green}} or {\color{Red} \dotuline{Red}}. Since the latter is more likely, he proceeds from $v_0$ to $v_1$. Overall, he wins with probability $\Pr({\color{blue} \text{\dashuline{Blue}}}) + \Pr({\color{red} \text{\dotuline{Red}}}) = \frac{5}{6}$. 
\end{example}

\paragraph*{Definitions}
We define reachability {\em probabilistic CRA} (p-CRA) games. A p-CRA game $\G = \zug{V, E, l, T, \Pr}$ is a reachability CRA game that is equipped with a probability distribution over colors $\Pr:[k]\to[0,1]$, i.e., $\sum_{g \in [k]} \Pr(g)=1$, where $\Pr[g]$ intuitively represents the probability that \PT's true color is $g$. For a \PO strategy $\sigma_1$ and a color $g \in [k]$, let $\play(v_0, \sigma_1, g)$ be the play in which \PO follows $\sigma_1$ and \PT follows the strategy that at $v \in V_2$, chooses the edge labeled $g$ (recall that in CRAs, there is only one such edge). For a \PO strategy $\sigma_1$, let $W[\sigma_1] \subseteq [k]$ be the colors that $\sigma_1$ wins against, i.e., $g \in W[\sigma_1]$ iff \PO wins $\play(v_0, \sigma_1, g)$. The {\em value} of $g$, denoted $\val(\sigma_1)$, is the probability that it wins, namely $\val(\sigma_1) = \sum_{g \in W[\sigma_1]} \Pr[g]$. We seek a \PO strategy with the maximal value.

\stam{
OLD

In the beginning of the game, Player 2 chooses a color $i$ by casting a $k$ sided dice with probabilities assigned by $\Pr$. Player 1 does not know which color Player 2 chose.
Player 2 deletes all the edges that disagree with the chosen color and plays on the remaining graph. Formally, the game is played on the graph $\zug{V, E\setminus \{e\in E_2:l(e)\neq i\}}$.

We define the {\em reward of play} $\pi=play(c_0, \sigma_1, \sigma_2)$ on $\G$ to be $rew(\pi)=
\begin{cases}
    1,& \text{player 1 wins}\\
    0, & \text{otherwise}
\end{cases}$
We define $rew_{\sigma_1}(c_0)$, the {\em reward of a strategy} $\sigma_1$ from the configuration $c_0\in C$, when $\sigma_1$ is a Player 1 strategy, to be the probability that player 1 will win from $c_0$ when playing according to $\sigma_1$.
We observe that $rew_{\sigma_1}(c_0)=\sum\limits_{i\in[k]}\Pr(i)rew(play(c_0, \sigma_1,[k]))$

We define $rew_{\sigma_1}(v_0)$, the {\em reward of a strategy} $\sigma_1$ from the vertex $v_0\in V$, when $\sigma_1$ is a Player 1 strategy, to be the probability that player 1 will win from $v_0$ when playing according to $\sigma_1$.
We observe that $rew_{\sigma_1}(v_0)=rew_{\sigma_1}(c_0)$ where $c_0=\zug{v_0,[k]}$.

We define $Val(v_0)$ {\em the value of the game} $\G$ from the starting vertex $v_0\in V$ to be the probability that player 1 will win from $v_0$ when playing optimally.\hfill\break
Formally, $Val(v_0)=\max\limits_{\sigma_1}(rew_{\sigma_1}(v_0))=\max\limits_{\sigma_1}(\sum\limits_{i\in[k]}\Pr(i)rew(play(\zug{v_0,i}, \sigma_1,i)))$.
}

\paragraph*{Finding an optimal strategy}
The following theorem reduces the problem of finding the value and an optimal strategy in a reachability p-CRA game to an {\em Integer Linear Program}~(ILP) (see below). 
This implies an NP algorithm for the {\em threshold problem}, namely deciding whether the value of a game is at least a given constant, and provides a practical algorithm for finding values using ILP solvers, e.g.,~\cite{gurobi}.

\begin{theorem}
Finding the value of a reachability p-CRA game is reducible to an ILP of polynomial size. 
\end{theorem}
\begin{proof}
An ILP (see details, e.g., in~\cite{Sch99}) is similar to a Linear Program (LP) except that variables are constrained to be in $\set{0,1}$; a program is defined over a set of variables $X$, it includes linear constraints and a linear maximization objective, both over $X$. 
Consider a reachability p-CRA game $\G = \zug{V, E, l, T, \Pr}$ and an initial vertex $v_0$. Intuitively, an optimal assignment to the variables in the ILP will give rise to a \PO strategy $\sigma_1$. 
Consider a color $g \in [k]$ and let $\play(v_0, \sigma_1, g) = c_0, c_1,\ldots$ with $c_j = \zug{v_j, L_j}$, for $j \geq 0$. We assume that targets are sinks and consider plays of length at most $|V| \cdot k$, which suffices since we a winning play can be assumed not to loop in a copy and traverse at most $k$ different copies.  
\stam{
Consider a reachability p-CRA game $\G = \zug{V, E, l, T, \Pr}$ and an initial vertex $v_0$. 
We describe the intuition of the ILP construction. An optimal assignment to the variables will give rise to a \PO strategy $\sigma_1$. 
Consider a color $g \in [k]$ and let $\play(v_0, \sigma_1, g) = c_0, c_1,\ldots$ with $c_j = \zug{v_j, L_j}$, for $j \geq 0$. We assume that targets are sinks and consider plays of length at most $|V| \cdot k$, since we have $|V|$ vertices in a copy and a path visits at most $k$ different copies. 
}
We introduce variables $x_{i,g,u}$, for $1 \leq i \leq k\cdot |V|$, $g \in [k]$, and $u \in V$. The constraints ensure that $x_{i,g,u} = 1$ iff the play visits $u$ at time $i$, i.e., $u = v_i$. 
The program thus identifies which colors $\sigma_1$ wins against, namely $\sigma_1$ wins against $g$ if there is a time $i$ and a target $u \in T$ such that $x_{i,g,u} =1$. 
The objective is to maximize the sum of the winning colors weighed by their probability. 

We describe the constraints. All our variables take values in $\set{0,1}$. 

\begin{itemize}
\item The token is initially located on $v_0$: $\forall g\in[k]: x_{0,g,v_0}=1$

\item For every color $g \in [k]$ and time $i \in |V| \cdot k$, the token is located on at most one vertex (once the token reaches a sink, it will not be placed on the arena):
\[\forall g\in[k],i\in[kn]:  \sum\limits_{v\in V}x_{i,g,v}\leq 1\]

\item The play might terminate if the token reaches a sink and the following constraint ensures that once that happens, the token does not reappear: 
\[\forall g\in[k],i\in[kn-1]:  \sum\limits_{v\in V}x_{i,g,v}\geq \sum\limits_{v\in V}x_{i+1,g,v}\]

\item We require the token movements to be consistent with the edges. If the token is on a \PO vertex $v \in V_1$, it can only move to $u \in  V$ if there is an edge between them: 
\[\forall i\in[nk-1],g\in[k],v\in V_1,u\in V \text{ s.t. } \zug{v,u} \notin E: x_{i,g,v}+x_{i+1,g,u}-1\leq 0 \]
Indeed, if both $x_{i,g,v}$ and $x_{i+1,g,u}$ are assigned $1$ and there is no edge between them, then the inequality is violated. 

If the token is on a \PT vertex $v \in V_2$, we require that the edge to $u \in V$ is colored $g$:
\[\forall i\in[nk-1],g\in[k],v\in V_2,u\in V \text{ s.t. } \zug{v,u} \notin E \text{ or } g \notin l(\zug{v,u}):  x_{i,g,v}+x_{i+1,g,u}-1\leq 0\]

\item  Consider two colors $g_1,g_2\in[k]$ such that their plays agree until time $i$, thus \PO cannot distinguish between \PT's actions. Assume that at time $i$, the token is placed on a \PO vertex. We enforce that \PO chooses an action independent of the color. We introduce additional {\em agreement} variables:
\[\forall i\in[nk], g_1, g_2\in[k], v\in V: a_{i,g_1,g_2,v}\in \set{0,1} \]
The following constraints ensure that $x_{j,g_1,v}=x_{j,g_2,v}=1$ iff $a_{j, g_1, g_2, v} = 1$.
\[\forall i\in[nk], g_1, g_2\in[k], v\in V:
\begin{cases}
    x_{i,g_1,v}+x_{i,g_2,v}\leq 1+ a_{i,g_1,g_2,v}\\
    x_{i,g_1,v} \geq a_{i,g_1,g_2,v}\\
    x_{i,g_2,v} \geq a_{i,g_1,g_2,v}
\end{cases}\]
The first constraint ensures that if both $x_{i,g_1,v}$ and $x_{i,g_2,v}$ are assigned $1$, then so does $a_{i,g_1,g_2,v}$. The other direction is ensured by the other two constraints, namely if $a_{i,g_1,g_2,v}$ is assigned $1$, then respectively $x_{i,g_1,v}$ and $x_{i,g_2,v}$ must be assigned $1$.

For all $i\in[nk]$ colors $g_1, g_2\in[k]$, and \PO vertex $v\in V_1$ that is not a sink:
\[(( \sum\limits_{0\leq j<i,u\in V}a_{j,g_1,g_2,u})+ x_{i,g_1,v}+x_{i,g_2,v})-i\leq \sum\limits_{w\in V}a_{i+1,g_1,g_2,w}\]
\end{itemize}

We claim that if the plays agree up to time $i-1$, and both end at $v$ at time $i$, then the left-hand side evaluates to $1$. Indeed, by the above, since the paths agree upto time $i-1$, there is exactly one agreement variable in each of these time steps that is assigned $1$, and in addition, both $x_{i,g_1,v}$ and $x_{i,g_2,v}$ are assigned $1$ for a total of $i+1$, from which $i$ is subtracted. This means that the right-hand side must be allocated $1$, which, again, by the above, means that the token moves to the same vertex in both plays.

We argue that an assignment to the variables that satisfies these constraints gives rise to a \PO strategy $\sigma_1$. Suppose that $\G$ reaches 
a \PO configuration $\zug{v, L}$ at time $i$, i.e., $v \in V_1$, let $g', g \in L$. Observe that the last constraint ensures that if $x_{i+1, g, u}$ and $x_{i+1, g', u'}$ are assigned $1$, then $u = u'$. This means that it is well defined to set $\sigma_1(\zug{v, L}) = u$ such that $u$ is a neighbor of $v$ and there exists $g \in L$ such that $x_{i+1, g, u}$ is assigned $1$.
Finally, the maximization objective is: 
\[\max \sum\limits_{g\in[k]}(\Pr(g)(\sum\limits_{i\in[nk],v\in T}x_{i,g,v}))\]
We stress that for $g \in [k]$, the probability $\Pr(g)$ is a constant and part of the input. 
Observe that since $\play(v_0, \sigma_1, g)$ visits $v \in V$ at time $i$ iff $x_{i,g,v} = 1$, the play is winning for \PO iff there is a time point at which it visits a target. Thus, the maximization coincides with the value of $\sigma_1$. 
\end{proof}

\begin{remark}[Lower bound]
Note that the NP-hardness lower bound from CRAs (Lem.~\ref{lem:UCRA-NP-hard}) applies to p-CRAs. Given a p-CRA $\G$ and $c \in [0,1]$, we argue that it is NP-hard to decide whether there exists a \PO strategy $\sigma_1$ with $\val(\sigma_1) \geq c$. 
Consider the game constructed in the proof of (Lem.~\ref{lem:UCRA-NP-hard}), if we view it as a p-CRA game, regardless of the assigned probabilities, the value of the game is $\val(\sigma_1) = 1$ iff $\varphi$ is satisfiable. That is, if $\varphi$ is not satisfiable, then $\val(\sigma_1) < 1$.
We use this property to construct a similar game by adding the additional vertex $s_0\in V_2$ and color $k+1$, such that $out(s_0)=\{s_1,s\}$, the color $k+1$ is assigned the probability $\Pr(k+1)=1-c$ and the colors of the new edges are $l(\zug{s_0,s_1})=[k]$ and $l(\zug{s_0,s})=\{k+1\}$. The probability of reaching the vertex $s_1$ in this new game is $c$. Therefor, the value of the game is $\val(\sigma_1) = c$ iff $\varphi$ is satisfiable, and $\val(\sigma_1) < c$ if $\varphi$ is not satisfiable.
\end{remark}

\section{Playing with Mixed Strategies}
So far, we considered only {\em pure} strategies for \PO, namely strategies that move the token deterministically. In this section, we allow \PO to follow a {\em mixed strategy} that moves the token randomly. We focus on CRA games. The following example demonstrates the model.

\begin{example}\label{ex:inner-v3}
Consider the reachability game depicted in Fig.~\ref{fig:v3colorblind}. Edges that are not drawn lead to a losing sink. 
Note that \PO cannot guarantee winning. Indeed, if he proceeds to $u_6$ or $u_8$, \PT chooses the color $1$, and if he proceeds to $u_7$, \PT chooses the color $2$. 
Table~\ref{table:v3} summarizes all outcomes;
once \PO chooses a successor vertex (column) and \PT chooses a color (row), the table marks whether \PO wins (1) or loses (0). 

A mixed strategy chooses a probability distribution over columns. \PO applies worst-case reasoning: for any fixed choice, \PT will respond by choosing the worst row. For example, consider the uniform strategy that chooses $u_6, u_7$, and $u_8$ each with probability $1/3$. \PT responds by choosing the color $1$ so that \PO wins with probability only $1/3$. However, if \PO assigns a probability of $1/2$ to $u_7$ and $1/2$ to $u_8$, he guarantees winning with probability $1/2$ (an alternatively optimal strategy assigns probability $1/4$ to each of $u_6$ and $u_8$). Note that this is a lower bound on the probability of winning; indeed, if \PT chooses color the color $6$, \PO wins with probability $1$. 
\end{example}

\begin{figure}[htbp]
\centering
\begin{minipage}[b]{0.45\textwidth}
    \centering
    \resizebox{\linewidth}{!}{%
    \begin{tikzpicture}[
        roundnode/.style={circle, draw=black!60, fill=white!5, very thick, minimum size=4mm},
        smallround/.style={circle, draw=black!60, fill=white!5, very thick, minimum size=3.5mm},
        squarednode/.style={rectangle, draw=black!60, fill=white!5, very thick, minimum size=4mm}
    ]
        \draw (-3,0) node[roundnode] (v1) {$v_1$};
        \draw (0,1.25) node[roundnode] (v2) {$v_2$};
        \draw (0,-1.25) node[roundnode] (v3) {$v_3$};
        \draw (-2,1.25) node[squarednode] (u1) {$u_1$};
        \draw (-2,-1.25) node[squarednode] (u2) {$u_2$};
        \draw (1,2) node[squarednode] (u3) {$u_3$};
        \draw (1,1.25) node[squarednode] (u4) {$u_4$};
        \draw (1,0.5) node[squarednode] (u5) {$u_5$};
        \draw (1,-0.5) node[squarednode] (u6) {$u_6$};
        \draw (1,-1.25) node[squarednode] (u7) {$u_7$};
        \draw (1,-2) node[squarednode] (u8) {$u_8$};

        \draw (2.9,1.25) node[roundnode, double] (t1) {};
        \draw (2.9,-1.25) node[roundnode, double] (t2) {};

        \draw (v1) edge[->,draw=black!60, very thick] (u1);
        \draw (v1) edge[->,draw=black!60, very thick] (u2);
        
        \draw (v2) edge[->,draw=black!60, very thick] (u3);
        \draw (v2) edge[->,draw=black!60, very thick] (u4);
        \draw (v2) edge[->,draw=black!60, very thick] (u5);
        
        \draw (v3) edge[->,draw=black!60, very thick] (u6);
        \draw (v3) edge[->,draw=black!60, very thick] (u7);
        \draw (v3) edge[->,draw=black!60, very thick] (u8);

        \draw (u1) edge[->,draw=black, very thick] node[xshift=-1mm, yshift=1.5mm] {\scriptsize $\{3,4,5\}$} (v2);
        \draw (u2) edge[->,draw=black, very thick] node[xshift=-1mm, yshift=-1.5mm] {\scriptsize $\{4,5,6\}$} (v3);
        \draw (u2) edge[->,draw=black, very thick] node[rotate=50, xshift=-6mm, yshift=1.5mm] {\scriptsize $\{1,2,5\}$} (v2);
        \draw (u1) edge[->,draw=black, very thick] node[rotate=-50, xshift=-6mm, yshift=1.5mm] {\scriptsize $\{1,2,6\}$} (v3);

        \draw (u4) edge[->,draw=black, very thick] node[xshift=-1mm, yshift=1.5mm] {\scriptsize $\{1,2,3,4\}$} (t1);
        \draw (u7) edge[draw=black, very thick] node[xshift=-1mm, yshift=1.5mm] {\scriptsize $\{1,4,6\}$} (t2);
        \draw (u3) edge[bend left, draw=black, very thick] node[rotate=-15, xshift=-1mm, yshift=1.5mm] {\scriptsize $\{2,3,5\}$} (t1);
        \draw (u6) edge[bend left,->,draw=black, very thick] node[rotate=-15, xshift=-1mm, yshift=1.5mm] {\scriptsize $\{2,3,6\}$} (t2);
        \draw (u5) edge[bend right,->,draw=black, very thick] node[rotate=15, xshift=-1mm, yshift=2mm] {\scriptsize $\{1,4,5\}$} (t1);
        \draw (u8) edge[bend right,->,draw=black, very thick] node[rotate=15, xshift=-1mm, yshift=2mm] {\scriptsize $\{2,3,6\}$} (t2);
    \end{tikzpicture}%
    }
    \caption{A mixed strategy CRA reachability game.}
    \label{fig:fullcolorblind}
\end{minipage}%
\hfill
\begin{minipage}[b]{0.25\textwidth}
    \centering
    \resizebox{\linewidth}{!}{%
    \begin{tikzpicture}[
        roundnode/.style={circle, draw=black!60, fill=white!5, very thick, minimum size=5mm},
        smallround/.style={circle, draw=black!60, fill=white!5, very thick, minimum size=3.5mm},
        squarednode/.style={rectangle, draw=black!60, fill=white!5, very thick, minimum size=5mm}
    ]
        \draw (-1,0) node[roundnode] (v2) {$v_2$};
        \draw (0,0.75) node[squarednode] (u3) {$u_3$};
        \draw (0,0) node[squarednode] (u4) {$u_4$};
        \draw (0,-0.75) node[squarednode] (u5) {$u_5$};
        \draw (1.35,0) node[roundnode, double] (t1) {};

        \draw (v2) edge[->,draw=black!60, very thick] (u3);
        \draw (v2) edge[->,draw=black!60, very thick] (u4);
        \draw (v2) edge[->,draw=black!60, very thick] (u5);

        \draw (u4) edge[->, draw=black, very thick] node[xshift=-0.6mm, yshift=1.5mm] {\scriptsize $\{1,2\}$} (t1);
        \draw (u3) edge[bend left,->,draw=black, very thick] node[rotate=-15, xshift=-1mm, yshift=1.5mm] {\scriptsize $\{2,5\}$} (t1);
        \draw (u5) edge[bend right,->,draw=black, very thick] node[rotate=15, xshift=-1mm, yshift=1.5mm] {\scriptsize $\{1,5\}$} (t1);
    \end{tikzpicture}%
    }
    \caption{The $\{1,2,5\}$ copy of the game from $v_2$.}
    \label{fig:v2colorblind}
\end{minipage}%
\hfill
\begin{minipage}[b]{0.25\textwidth}
    \centering
    \resizebox{\linewidth}{!}{%
    \begin{tikzpicture}[
        roundnode/.style={circle, draw=black!60, fill=white!5, very thick, minimum size=5mm},
        smallround/.style={circle, draw=black!60, fill=white!5, very thick, minimum size=3.5mm},
        squarednode/.style={rectangle, draw=black!60, fill=white!5, very thick, minimum size=5mm}
    ]
        \draw (-1,0) node[roundnode] (v3) {$v_3$};
        \draw (0,0.75) node[squarednode] (u6) {$u_6$};
        \draw (0,0) node[squarednode] (u7) {$u_7$};
        \draw (0,-0.75) node[squarednode] (u8) {$u_8$};
        \draw (1.33,0) node[roundnode, double] (t1) {};

        \draw (v3) edge[->,draw=black!60, very thick] (u6);
        \draw (v3) edge[->,draw=black!60, very thick] (u7);
        \draw (v3) edge[->,draw=black!60, very thick] (u8);

        \draw (u7) edge[->, draw=black, very thick] node[xshift=-0.6mm, yshift=1.5mm] {\scriptsize $\{1,6\}$} (t1);
        \draw (u6) edge[bend left,->,draw=black, very thick] node[rotate=-15, xshift=-1mm, yshift=1.5mm] {\scriptsize $\{2,6\}$} (t1);
        \draw (u8) edge[bend right,->,draw=black, very thick] node[rotate=15, xshift=-1mm, yshift=1.5mm] {\scriptsize $\{1,6\}$} (t1);
    \end{tikzpicture}%
    }
    \caption{The $\{1,2,6\}$ copy of the game from $v_3$.}
    \label{fig:v3colorblind}
\end{minipage}
\end{figure}

\begin{table}[htbp]
    \centering
       	
\begin{minipage}{0.3\textwidth}
\centering
\begin{tabular}{ c | c c c }
& ${\color{black!50}\frac{1}{2}}$ & ${\color{black!50}\frac{1}{2}}$ \\
 $v_1$ & $u_1$ & $u_2$ \\ 
 \hline
 $1$ & $\frac{1}{2}$ & $\frac{2}{3}$ & ${\color{black!50}\frac{7}{12}}$\\  
 $2$ & $\frac{1}{2}$ & $\frac{2}{3}$ & ${\color{black!50}\frac{7}{12}}$\\
 $3$ & $\frac{2}{3}$ & $\frac{1}{2}$& ${\color{black!50}\frac{7}{12}}$ \\  
 $4$ & $\frac{2}{3}$ & $\frac{1}{2}$ & ${\color{black!50}\frac{7}{12}}$\\
 $5$ & $\frac{2}{3}$ & $\frac{2}{3}$ & ${\color{black!50}\frac{2}{3}}$\\
 $6$ & $1$ & $1$ & ${\color{black!50}1}$\\
\end{tabular}
\caption{The matrix game for the main copy $\tilde{G_{[k]}}$ from $v_1$.}
\label{table:v1}
\end{minipage}
\hspace{0.3cm}
\begin{minipage}{0.3\textwidth}
\begin{tabular}{ c | c c c c }
&{\color{black!50}$\frac{1}{3}$} & {\color{black!50}$\frac{1}{3}$} & {\color{black!50}$\frac{1}{3}$ }& \\
 $v_2$ & $u_3$ & $u_4$ & $u_5$ \\ 
 \hline 
 $1$ & $0$ & $1$ & $1$ & {\color{black!50}$\frac{2}{3}$}\\
 $2$ & $1$ & $1$ & $0$ & {\color{black!50}$\frac{2}{3}$}\\ 
 $5$ & $1$ & $0$ & $1$ & {\color{black!50}$\frac{2}{3}$}\\
\end{tabular}
\caption{The matrix game for the $\{1,2,5\}$ copy of the game from $v_2$.}
\label{table:v2}
\end{minipage}
\hspace{0.3cm}
\begin{minipage}{0.3\textwidth}
\centering
\begin{tabular}{ c | c c c c }
&{\color{black!50}$0$} & ${\color{black!50}\frac{1}{2}}$ & ${\color{black!50}\frac{1}{2}}$ & \\
 $v_3$ & $u_6$ & $u_7$ & $u_8$ \\ 
 \hline
 $1$ & $0$ & $1$ & $0$ & {\color{black!50}$\frac{1}{2}$}\\  
 $2$ & $1$ & $0$ & $1$ & {\color{black!50}$\frac{1}{2}$}\\
 $6$ & $1$ & $1$ & $1$& {\color{black!50}$1$} \\
\end{tabular}
\caption{The matrix game for the $\{1,2,6\}$ copy of the game from $v_3$.}
\label{table:v3}
\end{minipage}
\end{table}

\paragraph*{Definitions}
A {\em mixed strategy} for \PO is a function $\sigma_1:V^*\cdot V_1 \to [0,1]^V$ that, given a history of play, assigns a probability distribution over successor vertices; namely, for each finite path $\pi \in V^*$ that ends in $v$, we require $\sum_{u \in V} \sigma_1(\pi)(u) = 1$ and only neighbors of $v$ are assigned positive probability, i.e., $\sigma_1(\pi)(u) > 0$ implies $u \in N(v)$.


Consider an initial configuration $c_0 = \zug{v_0, L}$. A mixed \PO strategy $\sigma_1$ and a \PT color $g \in L$ induce a probability distribution over infinite plays denoted $\Pr^{c_0,\sigma_1, g}$. For brevity, we omit $c_0,\sigma_1$, and $g$ when they are clear from the context. 
Consider a play $\pi = c_0, c_1, \ldots$. 
We define the probability of finite prefixes of $\pi$. 
For $n \in \Nat$, we use $\pi^{\leq n}$ to denote the prefix of length $n$ of $\pi$, thus $\pi^{\leq n} = c_0,\ldots, c_n$. For $0 \leq j \leq n$, let $c_j = \zug{v_j, L_j}$. 
We say that $\pi^{\leq n}$ is {\em consistent} with $\sigma_1$ and $g$ if for every $0 \leq i \leq n$, if $v_i \in V_1$, then $\sigma_1(\pi^{\leq n})(v_{i+1}) >0$, and if $v_i \in V_2$, then $g \in l(v_i, v_{i+1})$. 
We say that $\pi$ is consistent with $\sigma_1$ and $g$ if all its prefixes are consistent with the strategies. 
If $\pi^{\leq n}$ is not consistent with either strategy, we define $\Pr[\pi^{\leq n}] = 0$. 
Otherwise, we define $\Pr[\pi^{\leq n}] = \prod_{0 \leq i \leq n \text{ s.t. } v_i \in V_1} \sigma_1(\pi^{\leq i})(v_{i+1})$. 
We have defined a pre-measure over the set of infinite paths and can extend it to a measure by applying Carathéodory’s extension theorem~\cite{Bil95}. We note, however, that the \PO strategies that we will construct eventually choose actions deterministically, which implies a simpler definition. For an infinite play $\pi = v_0, v_1,\ldots$ that is consistent with $\sigma_1$ and $g$, let $n_0 \in \Nat$ be the index after which $\sigma_1$ chooses pure actions, then $\Pr[\pi] = \Pr[\pi^{\leq n_0}]$. 

\begin{definition}
The {\em value} of a strategy $\sigma_1$ from $c_0 = \zug{v_0, L}$ is $\val(\sigma_1, c_0) = \min_{g \in L} \Pr^{c_0,\sigma_1, g}[\set{\pi: \pi \models p}]$. The value of $c_0$ is $\val(c_0) = \max_{\sigma_1} \min_{g \in L} \Pr^{c_0,\sigma_1, g}[\set{\pi: \pi \models p}]$. An {\em optimal mixed strategy} maximizes the expression. 
\end{definition}

\paragraph*{Finding an optimal mixed strategy}
The following example illustrates our algorithm.

\begin{example}
Consider the game depicted in Fig.~\ref{fig:fullcolorblind}.
The algorithm works bottom up. It first solves the sub-games whose roots are $v_2$ and $v_3$.
Note that there are two paths that lead to each of these vertices and the paths do not fully agree on the colors, thus two copies of each sub-game need to be solved, and we depict only one of these copies for each game in Figs.~\ref{fig:v2colorblind} and~\ref{fig:v3colorblind}. Example~\ref{ex:inner-v3} describes the solution to the game $v_3$. 
For $v_2$, Table~\ref{table:v2} summarizes the outcomes for each of the pairs of actions of the two players. Here, it is optimal for \PO to choose a successor vertex (column) uniformly at random, since then, no matter which color (row) \PT chooses, \PO wins with probability $2/3$. 


Having solved the inner games and identified the value for each color, we proceed to solve the ``outer'' game at $v_1$. We use our solutions for the inner games to form Table~\ref{table:v1}. For example, suppose that \PT's color is $1$. 
If \PO proceeds to $u_2$, \PT proceeds to $v_2$, and the uniform strategy in the inner game guarantees winning with probability $2/3$. We stress that when the token reaches $v_2$, \PO is not aware of \PT's true color since her choices are consistent with colors $1,2$, and $5$, thus in the inner game, randomization is required since she can still choose any of the three. 
If \PO proceeds to $u_1$, \PT moves the token across the edge to $v_3$ with the color $1$, and using \PO's previously constructed strategy, \PO wins with probability $1/2$. 
An optimal strategy for \PO in this matrix game is to choose uniformly at random between $u_1$ and $u_2$, which guarantees winning with probability $7/12$.
\end{example}

The following theorem develops an algorithm for finding an optimal strategy.

\begin{theorem}
\label{thm:mixed}
Given a parity CRA game $\G$ and an initial vertex $v_0\in V$, finding the value of the game from $v_0$ is in PSPACE.
\end{theorem}
\begin{proof}
Recall that for $L \subseteq [k]$, copy $\tG_L$ of the explicit game $\tG$, contains configurations in which \PT has so far played consistently with the colors in $L$. 
In App.~\ref{app:mixed}, we describe an algorithm that proceeds in a DFS traversal over copies, similar in spirit to the warm-up algorithm in Sec.~\ref{sec:NP-upper}, 
but the recursion now involves constructing and solving a two-player zero-sum {\em matrix game} for each copy. \PO's actions in the matrix game are vertices (columns) and \PT's actions are colors (rows). \PO's utility is given by a function $\util: V \times [k] \rightarrow \set{0,1}$, which the algorithm constructs recursively. 
\end{proof}

\section{Variants}
In this section, we introduce variations of games with color-restricted opponents. 

\subsection{Playing against an adversary that changes her mind}
In previous sections, the adversary chooses actions that are consistent with a single color. In this section, we consider adversaries that can change their color mid-game.
To motivate this model, we return to the warehouse robots described in Section~\ref{intro}. Recall that the goal there is to design a controller for Robot~$R_2$ to operate in a warehouse in which Robot~$R_1$ is already deployed, while assuming prior knowledge of how $R_1$ operates: once a package is requested, it follows the shortest path to collect it and delivers it to a deposit location. That is, colors correspond to package locations. Now, once a package is delivered, $R_1$ is called to a different package locations; that is, at the deposit location, the adversary can change its color. 

\paragraph{Definitions.}
A {\em color-changing adversary} (CCA) reachability game, is a CRA game $\G=\zug{V,E,k,l,T}$ with the modification that $l:E_2\to 2^{[k]}\cup\{reset\}$ and every \PT edge $e\in E_2$ is either associated with a set of colors $l(e)\subseteq[k]$ or with the label $l(e)=reset$. 
When the game crosses an edge labeled $reset$, \PT resets her possible colors to $[k]$. Formally, consider a configuration $c=\zug{v,L}\in C$ with $v \in V_2$. Suppose that an edge $e = \zug{v, v'} \in E_2$ is crossed, and denote the next configuration by $c'=\zug{v',L'}$. If $l(e) \neq reset$, then the update is as in previous sections, namely $L' = l(e) \cap L$, and if $l(e) = reset$, then $L' = [k]$. In other words, reset edges direct to the initial copy $\tilde{G}_{[k]}$ of the explicit game.

\begin{theorem}
    Deciding whether \PO has a winning strategy from an initial vertex in CCA reachability games is in NP.
\end{theorem}
\begin{proof}
Consider a reachability CCA game $\G=\zug{V,E,k,l,T}$. Assume wlog. that targets are sinks, thus a winning play crosses reset edges only finitely often. 
We iteratively compute a set of vertices $S_i \subseteq V$, for $i \geq 0$, from which \PO wins by crossing a reset edge at most $i$ times. 
First, the set $S_0\in V$ is computed by redirecting every reset edge to a sink $s$ that is winning for \PT and applying the algorithm to solve CRA games (Lem.~\ref{lem:UCRA-NP}) starting from every vertex $v \in V$. 
Next, for $i >0$, assume that $S_{i-1}$ is computed. To compute $S_i$, we construct a game $\tilde{G'}=\zug{C',A',T'}$ based on the explicit graph of $\G$. We add two new sinks $t$ and $s$ that are respectively winning for Players~$1$ and~$2$, and re-direct every reset edge $e = \zug{v, v'}$ to one of them: if $v' \in S_{i-1}$, then we redirect $e$ to $t$, and otherwise to $s$. 
We again apply to $\tilde{G'}$, the algorithm in Lem.~\ref{lem:UCRA-NP} from every vertex, and obtain $S_i$ by taking the union of $S_{i-1}$ and the winning vertices in the main copy. We continue until a fixed-point $S^* = S_i = S_{i-1}$ is reached. Observe that from every vertex $v \notin S^*$, \PT forces that either a copy that is losing for \PO is reached or a cycle using reset edges is formed. Both outcomes are losing for \PO. 
We adjust the algorithm to be in NP by guessing a \PO strategy once in the beginning of the algorithm and using it subsequently for the recursive applications of the algorithm in Lem.~\ref{lem:UCRA-NP}. The guess is polynomial since each \PO strategy used in the inner algorithm is polynomial, and the inner algorithm is called once for every vertex in every iteration, and there are at most $|V|$ iterations. 
\end{proof}


\subsection{A symmetric game}
We consider a variant in which both players are color restricted. We observe, somewhat surprisingly, a drop in complexity.

\paragraph{Definitions.}
A parity {\em symmetric color-restricted game} is $\G=\zug{V,E,\{k_1,k_2\},\{l_1,l_2\},p}$, where as before, the vertices are partitioned among the players $V = V_1 \cup V_2$, \PLi's edges are $E_i = \set{e = \zug{v, v'} \in E: v \in V_i}$, for $i \in \set{1,2}$, the parity function $p: V \rightarrow \set{0,\ldots, d}$ specifies \PO's objective, and both players are color restricted: for $i\in{1,2}$, \PLi's possible colors are $[k_i]$ and his labeling function is $l_i: E_i \rightarrow 2^{[k_i]}$. 
A configuration is now $c = \zug{v,L_1,L_2}$, where for $i \in \set{1,2}$, the set $L_i \subseteq [k_i]$ marks the colors that are consistent with \PLi's previous choices. When $v \in V_i$, \PLi chooses an outgoing edge $e = \zug{v, v'}$ from $v$, where $e$ is available for \PLi if $l(e)\in L_i$, and the next configuration is $\zug{v', L'_1, L'_2}$ with $L'_i = L_i \cap l(e)$ and the other set is not updated, namely $L'_j = L_j$, for $j \neq i$.

\begin{theorem}
    Deciding whether \PO wins a parity symmetric color-restricted game is in P.
\end{theorem}
\begin{proof}
Consider a game $\G$. Observe a configuration $c = \zug{v, L_1, L_2}$ within a copy of the corresponding explicit game $\tilde{G}_{L_1,L_2}=\zug{V_{L_1,L_2},E_{L_1,L_2}}$, where $L_i \subseteq [k_i]$, for $i \in \set{1,2}$. 
Suppose that \PO controls $c$, and the claim is dual for \PT. 
Observe that if $c$ does not have outgoing edges that exit its copy, then it has a unique outgoing edge $e$ with $L_i \subseteq l_1(e)$. 
Thus, from each configuration, there is a unique path in its copy, and the path leads either to: (1)~a sink, (2)~a simple loop, or (3)~a vertex with outgoing edges that exit the copy. In cases (1) and (2), we say that these configurations belong to a {\em sink}, and define that a sink is winning depending on whether the loop satisfies \PO's parity objective. Similarly to the graph $\tG[\sigma_1]$ in Lem.~\ref{lem:tree}, we can now conclude the explicit graph $\tilde{G}$ of $\G$ is a tree with cycles only
in its bottom strongly-connected components. A calculation as in Lem.~\ref{lem:small-copies-tree} shows that the number of reachable copies is at most $2 \cdot k_1 - 1 +k_1(2 \cdot k_2 - 1)$, thus the number of reachable configurations is polynomial. To decide whether \PO wins, we first apply a DFS algorithm to identify the reachable configurations, then solve the game in a backwards induction manner from the sinks upwards. 
\end{proof}


\subsection{Generalized CRA games}
{\em Generalized CRA games} allow multiple edges with the same color. 
We point out that our solution of the problem of deciding whether \PO wins a generalized CRA game coincides with the problem of deciding whether a multi-environment MDP {\em surely} satisfies an objective, but these have only been studied under {\em almost-sure} semantics. Our solution is presented in App.~\ref{app:general-CRA} for completeness. 

\begin{theorem}
\label{thm:general-CRA}
Deciding whether \PO wins a CRA reachability game is PSPACE-complete.
\end{theorem}

\section{Discussion}
\label{sec:disc}
We introduce and study CRA games with a restricted opponent. CRA games can model an environment wth an agent that given an objective follows a known algorithm to select a strategy, e.g., travel on the shortest path, but the objective that the agent is targeting is unknown. CRA games exhibit relatively efficient algorithms when compared with multi-environment MDPs, and their simplicity invites a study of technically challenging solution concepts like mixed strategies as well as extensions and variants. 
There are plenty of directions for future work. 
First, it is interesting to study games incorporating multiple color-restricted players. This could model, for example, a robot introduced to a warehouse with multiple robots. 
Second, a study of {\em repair} in CRA games (see, e.g.,~\cite{AAK15}) to address situations in which \PO does not win. For example, a way to address non-realizability of a controller for robot $R_2$ would be to recall the deployed $R_1$ and alter its strategies (colors). 
Third, CRA games are essentially games with a specific form of partial information, and this restriction leads to improved complexity results over general partial-information games. This is a broad concept. To demonstrate, a {\em bidding game}~\cite{AHC19} is a two-player graph game in which both players have budgets and an auction (bidding) determines which player moves in each turn. In real-life, participants of an auction are rarely fully-informed of their opponent's budget. This notion of partial information that is specific to bidding games is modeled in~\cite{AJZ23} by assuming that the players' budget are drawn from a known probability distribution. Studying further specific notions of partial information is an exciting direction for future work. 

\small

 \bibliographystyle{splncs04}
 \bibliography{ga}

\newpage
\appendix
\begin{figure}
	\centering
	\begin{tikzpicture}[
        roundnode/.style={circle, draw=black!60, fill=white!5, very thick, minimum size=5mm},smallround/.style={circle, draw=black!60, fill=white!5, very thick, minimum size=3.5mm},
        squarednode/.style={rectangle, draw=black!60, fill=white!5, very thick, minimum size=5mm},
        ]
		\draw (-5,0) node[roundnode] (0rb) {$v_0$};
		\draw (-4,0) node[squarednode] (1rb) {$v_1$};
        \draw (-6,1) node[squarednode] (2rb) {$v_2$};
		\draw (-6,-1) node[squarednode] (3rb) {$v_3$};
        \draw (-4,1) node[smallround] (4rb) {$v_4$};
        \draw (-4,-1) node[smallround] (5rb) {$v_5$};
        \draw (-7,1) node[roundnode] (srb) {$s$};
        \draw (-7,-1) node[roundnode, double] (trb) {$t$};

        \draw (0,0) node[roundnode] (0r) {$v_0$};
		\draw (1,0) node[squarednode] (1r) {$v_1$};
        \draw (-1,1) node[squarednode] (2r) {$v_2$};
		\draw (-1,-1) node[squarednode] (3r) {$v_3$};
        \draw (1,1) node[smallround] (4r) {$v_4$};
        \draw (1,-1) node[smallround] (5r) {$v_5$};
        \draw (-2,1) node[roundnode] (sr) {$s$};
        \draw (-2,-1) node[roundnode, double] (tr) {$t$};

        \draw (5,0) node[roundnode] (0b) {$v_0$};
		\draw (6,0) node[squarednode] (1b) {$v_1$};
        \draw (4,1) node[squarednode] (2b) {$v_2$};
		\draw (4,-1) node[squarednode] (3b) {$v_3$};
        \draw (6,1) node[smallround] (4b) {$v_4$};
        \draw (6,-1) node[smallround] (5b) {$v_5$};
        \draw (3,1) node[roundnode] (sb) {$s$};
        \draw (3,-1) node[roundnode, double] (tb) {$t$};

        \draw (-5,-1) edge[->,draw=black!60, very thick, minimum size=7mm] node[] {} (0rb);

        \filldraw[black] (-5,-1) circle (2pt) node[]{};

        \draw (-7.5,-1.5) rectangle (-3.5,1.5)[draw=red, loosely dotted, very thick];
        \draw (-7.5,-1.5) rectangle (-3.5,1.5)[draw=blue!70, loosely dashed,  thick];
        
        \draw (-2.5,-1.5) rectangle (1.5,1.5)[draw=red, dotted, very thick];
        \draw (2.5,-1.5) rectangle (6.6,1.5)[draw=blue, dashed, thick];

		\draw (0rb) edge[->,draw=black!60, very thick, minimum size=7mm] node[] {} (1rb);
        \draw (0rb) edge[->,draw=black!60, very thick, minimum size=7mm] node[] {} (2rb);
        \draw (0rb) edge[->,draw=black!60, very thick, minimum size=7mm] node[] {} (3rb);

        \draw (0r) edge[->,draw=black!60, very thick, minimum size=7mm] node[] {} (1r);
        \draw (0r) edge[->,draw=black!60, very thick, minimum size=7mm] node[] {} (2r);
        \draw (0r) edge[->,draw=black!60, very thick, minimum size=7mm] node[] {} (3r);

        \draw (0b) edge[->,draw=black!60, very thick, minimum size=7mm] node[] {} (1b);
        \draw (0b) edge[->,draw=black!60, very thick, minimum size=7mm] node[] {} (2b);
        \draw (0b) edge[->,draw=black!60, very thick, minimum size=7mm] node[] {} (3b);

        \draw (4rb) edge[->,draw=black!60, very thick, minimum size=7mm] node[] {} (0rb);
        \draw (5rb) edge[->,draw=black!60, very thick, minimum size=7mm] node[] {} (0rb);

        \draw (4r) edge[->,draw=black!60, very thick, minimum size=7mm] node[] {} (0r);
        \draw (5r) edge[->,draw=black!60, very thick, minimum size=7mm] node[] {} (0r);

        \draw (4b) edge[->,draw=black!60, very thick, minimum size=7mm] node[] {} (0b);
        \draw (5b) edge[->,draw=black!60, very thick, minimum size=7mm] node[] {} (0b);

        \draw (1rb) edge[out=-60, in=-130, looseness=0.7, ->,draw=red, very thick, dotted, minimum size=7mm] node[] {} (5r);
        \draw (1rb) edge[out=60, in=140, out looseness=1, in looseness=0, ->,draw=blue, very thick, dashed, minimum size=7mm] node[] {} (4b);

        \draw (1r) edge[->,draw=red, very thick, dotted, minimum size=7mm] node[] {} (5r);
        \draw (1b) edge[->,draw=blue, very thick, dashed, minimum size=7mm] node[] {} (4b);

        \draw (2rb) edge[out=15, in=140, looseness=0.5,->,draw=blue, very thick, dashed, minimum size=7mm] node[] {} (sb);
        \draw (2rb) edge[out=40, in=100, out looseness=0.7, in looseness=2, ->,draw=red, very thick, dotted, minimum size=7mm] node[] {} (tr);

        \draw (2b) edge[->,draw=blue, very thick, dashed, minimum size=7mm] node[] {} (sb);
        \draw (2r) edge[->,draw=red, dotted, very thick, minimum size=7mm] node[] {} (tr);

        \draw (3rb) edge[out=-40, in=-100, out looseness=0.7, in looseness=2, ->,draw=red, very thick, dotted, minimum size=7mm] node[] {} (sr);
        \draw (3rb) edge[out=-15, in=-140, looseness=0.6, ->,draw=blue, very thick, dashed, minimum size=7mm] node[] {} (tb);

        \draw (3r) edge[->,draw=red, very thick, dotted, minimum size=7mm] node[] {} (sr);
        \draw (3b) edge[->,draw=blue, very thick, dashed, minimum size=7mm] node[] {} (tb);

	\end{tikzpicture}

	\caption{The explicit game $\tG$ that corresponds to the game depicted in Fig.~\ref{fig:ex1}. A copy $L \subseteq \set{{\color{Blue} \text{\dashuline{Blue}}}, {\color{Red} \text{\dotuline{Red}}}}$ is depicted with a ``frame'' colored with the colors in $L$. Observe that the initial configuration is $\zug{v_0, \set{{\color{Blue} \text{\dashuline{Blue}}}, {\color{Red} \text{\dotuline{Red}}}}}$.}\label{fig:ex1exp}
\end{figure}

\section{Proof of Prop.~\ref{prop:memory-req}}
\label{app:memory-req}
We generalize the game that is depicted in Fig.~\ref{fig:ex1}. The initial vertex is $v_0 \in V_1$ from which \PO can either proceed left or right. On the left-hand side of the game, there are $k$ vertices $v_1, \ldots, v_k \in V_2$. For $i \in [k]$, the outgoing edge from $v_i$ colored $i$ directs to the target $t$ and all other colors lead to a losing sink $s$. Intuitively, before turning left, \PO must narrow down \PT's choice of color to a single color $i \in [k]$ and only then, proceed to $v_i$. Indeed, if \PT's previous choices are consistent with two colors $i$ and $j$, and \PO proceeds to $v_i$, then \PT wins by choosing Color~$j$ and proceeding to $s$. 

The right-hand side of the game (including $v_0$) consists of $k$ vertices. For each $0 \leq i < \log k$ and $1 \leq j \leq 2^i$, we have a vertex $u^j_i \in V_2$ whose edges either lead back to $v_0$ or to $s$ as we describe now. The vertex $u^1_0$ has two outgoing edges, labeled respectively with colors $\set{1,\ldots, k/2}$ and $\set{k/2+1,\ldots, k}$. Thus, when \PT's choice at $u^1_0$ rules out half the colors. \PT's action at vertices $u^1_1$ and $u^1_2$ rule out a further half of the colors; each of the two vertices has two outgoing edges to $v_0$ labeled by colors $\set{1,\ldots, k/4}$ and $\set{k/4+1,\ldots, k/2}$ for $u^1_1$ and  $\set{k/2+1,\ldots,3k/4}$ and $\set{3k/4+1,\ldots, k}$ for $u^2_1$. An  additional outgoing edge from each vertex with the remaining colors leads to $s$, thus \PO proceeds to $u^1_1$ and $u^2_1$ only after visiting $u^1_0$ and ruling out half the colors, otherwise \PT wins. The construction continues similarly: for $0 \leq i < \log k$ and $1 \leq j \leq 2^i$, the vertex $u^j_i$ is associated with a set of colors $L(u^j_i)$ that have not been ruled out yet, where $|L(u^j_i)| = k/2^i$, and \PT's action at $u^j_i$ rules out half the colors in $L(u^j_i)$. \PO then proceeds to corresponding vertex in the next level. Finally, \PT's choice at $u^j_{\log k -1}$ narrows down her color to a unique color.

We have already describe \PO's winning strategy. Observe that when he follows it, when \PT chooses a color in $L(u^j_i)$, then the play visits $u^j_i$ in the copy $L(u^j_i)$. Thus, $2k-1$ copies are reachable. 

\section{Proof of Thm.~\ref{thm:mixed}}
\label{app:mixed}
The base case of the algorithm is a singleton $L = \set{g}$. For each $v \in V$, define $\util(v, g) = 1$ iff there is a path that satisfies \PO's parity objective from $v$, where \PT is only allowed to traverse $g$-colored edges. This is clearly decidable in polynomial time. Moreover, observe that if $\util(v, g) = 1$, \PO can guarantee winning from $\zug{v, \set{g}}$ and if $\util(v, g) = 0$,  \PT can choose the color $g$ at $\zug{v, \set{g}}$ and violate \PO's objective. 

For the recursive step, consider $v \in V_1$ and $L \subseteq [k]$. First, if there is a path from $\zug{v, L}$ that satisfies the parity objective and is contained in $\tG_L$, \PO wins from $\zug{v, L}$ and we define $\util(v, g) = 1$, for every $g \in L$. Otherwise, we traverse copy $\tG_L$ in a BFS manner and collect a set of \PT vertices $O(v,L) \subseteq V_2$ that lead out of copy $L$, namely $u \in O(v, L)$ if there is a path in $\tG_L$ from $\zug{v, L}$ to configuration $\zug{u,L}$ and the neighbors of $\zug{u,L}$ reside in lower copies, i.e., $\zug{u', L'} \in N(\zug{u,L})$ implies $L' \subset L$. Note that since $\G$ is a CRA game, the neighbors of $\zug{u,L}$ form a partition of $L$. 
If $O(v, L) = \emptyset$, then \PO cannot win from $\zug{v, L}$ and we define $\util(v, g) = 0$, for every $g \in L$. Otherwise, we apply the recursive algorithm to each neighbor $\zug{u', L'}$, to obtain $\util(u', g)$, for every $g \in L'$. 

Assume by induction that, for each $c = \zug{u, L}$, we have $\val(c) = min_{g \in L} \util(u, g)$; namely, \PO has a strategy $\sigma_1^c$ that guarantees winning with probability at least $\val(c)$ and \PT has a choice of color $g \in L$ that achieves this probability. 
We propagate these values. 
Consider the matrix game in which \PO's actions are $O(v, L)$ and \PT's actions are $L$. For each choice of (pure) actions $u \in O(v, L)$ and $g \in L$, \PO's utility is $\util(v, g)$. It is well-known that finding \PO's max-min strategy $\sigma_1^L$ can be done in polynomial time via a reduction to linear programming~\cite{Dan66}. That is, $\sigma_1^L$ is a probability distribution over $O(v, L)$ that maximizes $\mathcal{V}(v, L) = \min_{g \in L} \sum_{u \in O(v, L)} \util(v, g)$. Observe first that a \PO strategy $\sigma_1$ that initially agrees with $\sigma^L_1$ by choosing the path to each $u \in O(v, L)$ with probability $\sigma^L_1(u)$ and then agrees with $\sigma_1^{\zug{u, L}}$ guarantees winning with probability $v(v, L)$. On the other hand, choosing the color $g \in L$ that minimizes the expression, attains a winning probability of $\mathcal{V}(v, L)$. Thus, $\mathcal{V}(v,L) = \val(\zug{v, L})$.

\section{Proof of Thm.~\ref{thm:general-CRA}}
\label{app:general-CRA}
\begin{figure}

	\centering

	\begin{tikzpicture}[
        roundnode/.style={circle, draw=black!60, fill=white!5, very thick, minimum size=5mm},smallround/.style={circle, draw=black!60, fill=white!5, very thick, minimum size=3.5mm},
        squarednode/.style={rectangle, draw=black!60, fill=white!5, very thick, minimum size=5mm}, tightzigzag/.style={decorate, decoration={zigzag, segment length=1.5mm, amplitude=.5mm}}
        ]

        \draw (-6,0) node[squarednode] (s11) {$s^1_1$};		
        \draw (-3,0) node[squarednode] (s12) {$s^1_2$};
        \draw (0,0) node[roundnode] (s21) {$s^2_1$};
        \draw (3,0) node[roundnode] (s22) {$s^2_2$};
        \draw (6,0) node[roundnode] (s) {$s$};

        \draw (-5,0.5) node[squarednode] (y1) {$y_1$};
        \draw (-5,-0.5) node[squarednode] (ny1) {$\overline{y_1}$};

        \draw (-2,0.5) node[squarednode] (y2) {$y_2$};
        \draw (-2,-0.5) node[squarednode] (ny2) {$\overline{y_2}$};

        \draw (1,0.5) node[squarednode] (x1) {$x_1$};
        \draw (1,-0.5) node[squarednode] (nx1) {$\overline{x_1}$};

        \draw (4,0.5) node[squarednode] (x2) {$x_2$};
        \draw (4,-0.5) node[squarednode] (nx2) {$\overline{x_2}$};

        \draw (-3,-2) node[roundnode, double, minimum size=7mm] (c1) {$C_1$};
        \draw (0,-2) node[roundnode, double, minimum size=7mm] (c2) {$C_2$};
        \draw (3,-2) node[roundnode, double, minimum size=7mm] (c3) {$C_3$};

        \draw (-6.75,0) edge[->,draw=black!60, very thick, minimum size=7mm] node[left] {} (s11);

        \filldraw[black] (-6.75,0) circle (2pt) node[]{};

		\draw (s21) edge[->,draw=black!60, very thick, minimum size=7mm] node[] {} (x1);
        \draw (s21) edge[->,draw=black!60, very thick, minimum size=7mm] node[] {} (nx1);
        
        \draw (s22) edge[->,draw=black!60, very thick, minimum size=7mm] node[] {} (x2);
        \draw (s22) edge[->,draw=black!60, very thick, minimum size=7mm] node[] {} (nx2);

        \draw (s11) edge[out=-45, in=-160, looseness=0.5,->,draw=red, very thick, dotted, minimum size=7mm] node[] {} (ny1);
        \draw (s11) edge[out=-30, in=-180, looseness=0.5,->,draw=blue, very thick, dashed, minimum size=7mm] node[] {} (ny1);
        \draw (s11) edge[out=-15, in=-210, looseness=0.5,->,draw=green,  thick, tightzigzag, minimum size=7mm] node[] {} (ny1);

        \draw (s11) edge[out=15, in=-160, looseness=0.5,->,draw=red, very thick, dotted, minimum size=7mm] node[] {} (y1);
        \draw (s11) edge[out=30, in=-180, looseness=0.5,->,draw=blue, dashed, very thick, minimum size=7mm] node[] {} (y1);
        \draw (s11) edge[out=45, in=-210, looseness=0.5,->,draw=green,  thick, tightzigzag, minimum size=7mm] node[] {} (y1);

        \draw (s12) edge[out=-45, in=-160, looseness=0.5,->,draw=red, very thick, dotted, minimum size=7mm] node[] {} (ny2);
        \draw (s12) edge[out=-30, in=-180, looseness=0.5,->,draw=blue, dashed, very thick, minimum size=7mm] node[] {} (ny2);
        \draw (s12) edge[out=-15, in=-200, looseness=0.5,->,draw=green, thick, tightzigzag, minimum size=7mm] node[] {} (ny2);

        \draw (s12) edge[out=15, in=-160, looseness=0.5,->,draw=red, very thick, dotted, minimum size=7mm] node[] {} (y2);
        \draw (s12) edge[out=30, in=-180, looseness=0.5,->,draw=blue, dashed, very thick, minimum size=7mm] node[] {} (y2);
        \draw (s12) edge[out=45, in=-210, looseness=0.5,->,draw=green,  thick, tightzigzag, minimum size=7mm] node[] {} (y2);

        \draw (y1) edge[out=0, in=-270, looseness=0.5,->,draw=red, dotted, very thick, minimum size=7mm] node[] {} (c1);
        \draw (ny1) edge[out=-10, in=-160, looseness=0.5,->,draw=red, dotted, very thick, minimum size=7mm] node[] {} (s12);

        \draw (y2) edge[out=10, in=-200, looseness=0.5, out looseness=0.4, ->,draw=red, very thick, dotted, minimum size=7mm] node[] {} (s21);
        \draw (ny2) edge[out=-10, in=-160, looseness=0.5, ->,draw=red, very thick, dotted, minimum size=7mm] node[] {} (s21);

        \draw (x1) edge[out=-30, in=15, in looseness=1.4, out looseness=1.3, ->,draw=red, very thick, dotted, minimum size=7mm] node[] {} (c1);
        \draw (nx1) edge[->,draw=red, very thick, dotted, minimum size=7mm] node[] {} (s22);

        \draw (x2) edge[->,draw=red, very thick, dotted, minimum size=7mm] node[] {} (s);
        \draw (nx2) edge[out=-20, in=13, in looseness=2, out looseness=0.6, ->,draw=red, very thick, dotted, minimum size=7mm] node[] {} (c1);

        \draw (y1) edge[out=10, in=-200, looseness=0.5, in looseness=2,->,draw=blue, very thick, dashed, minimum size=7mm] node[] {} (s12);
        \draw (ny1) edge[->,draw=blue, very thick, dashed, minimum size=7mm] node[] {} (c2);

        \draw (y2) edge[out=0, in=-270, looseness=0.5,->,draw=blue, very thick, dashed, minimum size=7mm] node[] {} (c2);
        \draw (ny2) edge[->,draw=blue, very thick, dashed, minimum size=7mm] node[] {} (s21);

        \draw (x1) edge[out=10, in=-200, looseness=0.5, ->,draw=blue, very thick, dashed, minimum size=7mm] node[] {} (s22);
        \draw (nx1) edge[out=-10, in=-160, looseness=0.5, ->,draw=blue, very thick, dashed, minimum size=7mm] node[] {} (s22);

        \draw (x2) edge[out=0, in=10, in looseness=3, out looseness=0.4, ->,draw=blue, dashed, very thick, minimum size=7mm] node[] {} (c2);
        \draw (nx2) edge[->,draw=blue, very thick, dashed, minimum size=7mm] node[] {} (s);

        \draw (y1) edge[->,draw=green,  thick, tightzigzag, minimum size=7mm] node[] {} (s12);
        \draw (ny1) edge[->,draw=green,  thick, tightzigzag, minimum size=7mm] node[] {} (s12);

        \draw (y2) edge[->,draw=green,  thick, tightzigzag, minimum size=7mm] node[] {} (s21);
        \draw (ny2) edge[->,draw=green,  thick, tightzigzag, minimum size=7mm] node[] {} (c3);

        \draw (x1) edge[->,draw=green,  thick, tightzigzag, minimum size=7mm] node[] {} (s22);
        \draw (nx1) edge[out=0, in=-270, looseness=0.5,->,draw=green,  thick, tightzigzag, minimum size=7mm] node[] {} (c3);

        \draw (x2) edge[out=10, in=-200, looseness=0.5,->,draw=green,  thick, tightzigzag, minimum size=7mm] node[] {} (s);
        \draw (nx2) edge[out=0, in=50, in looseness=1, out looseness=1, ->,draw=green,  thick, tightzigzag, minimum size=7mm] node[] {} (c3);

	\end{tikzpicture}

	\caption{The generalized CRA reachability game constructed from the formula $\varphi =\forall y\exists x (C_1 \wedge C_2 \wedge C_3)$ with $\color{red}{\dotuline{C_1}}\color{black}=(x_1\vee \overline{x_2}\vee y_1)$, $\color{blue}{\dashuline{C_2}}\color{black}=(x_2\vee \overline{y_1}\vee y_2)$, and $\color{green!60!black}{\uwave{C_3}}\color{black}=(\overline{x_1}\vee \overline{x_2}\vee\overline{y_2})$.}\label{fig:multoptpayoff2}
\end{figure}
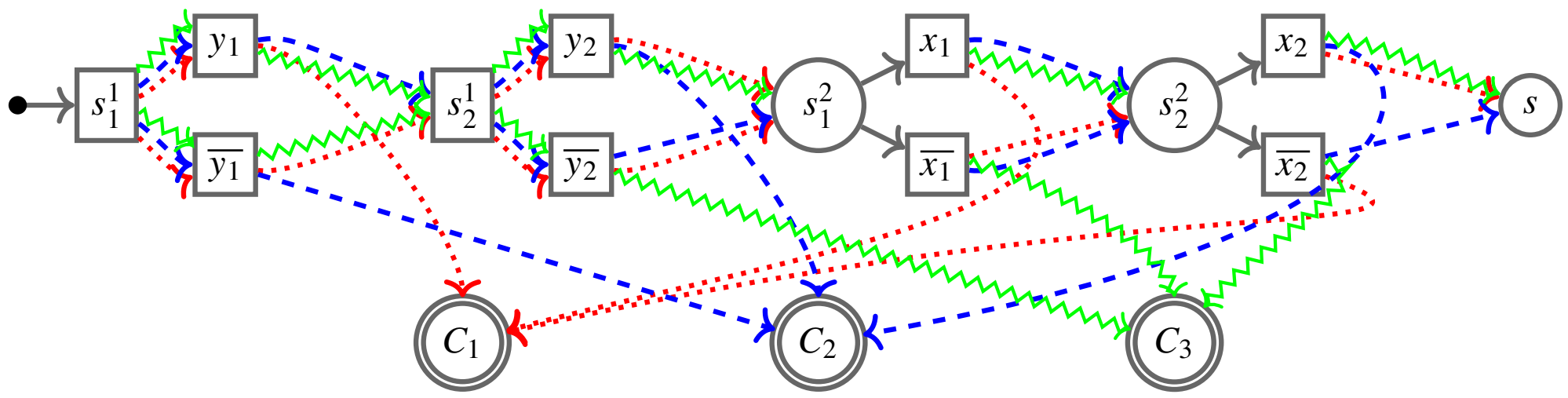

For the upper bound, we observe that the warm-up algorithm presented in Sec.~\ref{sec:NP-upper} applies to general CRA games and operates in PSPACE. 
For the lower bound, we show a reduction from TQBF. 
Consider a formula $\varphi=Q_1 X^i\dots Q_n X^n (C_1\wedge \ldots \wedge C_k)$ over variables $X^1\cup \ldots\cup X^n$, where for $1\leq i\leq n$, we have $X^i=\{x^i_1,\ldots,x^i_{m_i}\}$ and $Q_i\in\{\forall,\exists\}$.
We construct a generalized CRA reachability game $\G = \zug{V, E, k, l ,T}$ as follows. 
For $1\leq i\leq n$, we construct a gadget similar to the one constructed in Lem.~\ref{lem:UCRA-NP-hard} for each $X^i$ only that if $Q_i=\forall$, the vertices $s^i_j$ are \PT vertices, and all of their outgoing edges are associated with all the colors $C_1,\ldots,C_k$. That is, \PT chooses an assignment for the variables $s^i_j$. 
For example, see vertices $s^1_1, s^1_2$ and their outgoing edges in Fig.~\ref{fig:multoptpayoff}. We point out that we rely on general CRAs here; indeed, these \PT vertices have two outgoing edges labeled by the same color, thus choosing one of the edges does not rule out the color. 
Next, we connect the gadgets as follows. For each $1\leq i<n$, the outgoing edges from a literal $\ell^i \in \set{x^i_{m_i},\overline{x^i_{m_i}}}$ continues right and leads to $s^{i+1}_1$, the first vertex of the next gadget. 
For example, in Fig.~\ref{fig:multoptpayoff2}, the edges leaving $y_2,\overline{y_2}$ lead to $s^2_1$. 
The targets are sinks $\set{C_1,\ldots, C_k}$ that are shared by all gadgets. 

Formally, let $V_1 = \set{s^i_1,\ldots, s^i_{m_i}:1\leq i\leq n, Q_i=\exists}$ and $V_2 = \set{x^i_j, \overline{x^i_j}: 1 \leq i \leq n,1\leq j\leq m_i}\cup\set{s^i_j: 1 \leq i \leq n,1\leq j\leq m_i,Q_i=\forall}$. \PO's targets are $T = \set{C_1,\ldots, C_k}$.  
For $1 \leq i \leq n, 1\leq j< m_i$, the neighbors of $s^i_j$ are $x^i_j$ and $\overline{x^i_j}$. Each clause is associated with a color. If $Q_i=\forall$, than the edges connecting them are associated with all the colors, i.e. $l(\zug{s^i_j,\ell^i_j})=\set{C_1,\ldots,C_k}$ for $\ell^i_j \in \set{x^i_j, \overline{x^i_j}}$.
For $1 \leq i \leq n, 1\leq j< m_i$,  $\ell^i_j \in \set{x^i_j, \overline{x^i_j}}$, and clause $C_h,h\in[k]$, we define that the $C_h$-labeled edge from $\ell^i_j$ proceeds to target vertex $C_h \in T$ iff $\ell^i_j$ appears in $C_h$, and otherwise it proceeds to $s_{i+1}$ -- or to $s^{i+1}_1$ in the case that $j=m_i$, and $s$ in the case that $i=n, j=m_n$. 
Vertex $s$ is a sink and \PT wins if the game reaches $s$. 

Correctness is similar to Lem.~\ref{lem:UCRA-NP-hard}. A \PO strategy $\sigma_1$ in $\G$ corresponds to a {\em strategy} $f$ that chooses an assignment to the existentially-quantified variables based on the observed assignment to the universally-quantified variables. Dually, a \PT strategy $\sigma_2$ in $\G$ corresponds to {\em strategy} $g$ for the universally-quantified variables. If $\varphi$ is true, such an $f$ exists and guarantees that for every $g$, all clauses are satisfied. By following the corresponding $\sigma_1$, \PO guarantees that no matter which color (clause) \PT follows, the play will eventually lead to a target. Dually, if $\varphi$ is false, there is $g$ that guarantees that for every $f$, there is at least one clause that is not satisfied. By following the corresponding $\sigma_2$, \PT guarantees that there always is a color that is consistent with her choices that leads right, i.e., a clause that has not yet been satisfied by the joint partial assignment to variables.

\end{document}